\documentclass{aa}  

\usepackage{graphicx}
\usepackage{txfonts}

\usepackage{booktabs} 
\usepackage{amsmath}
\usepackage{placeins} 

\newcommand{\teff}{$T_{\mathrm{eff}}$}
\newcommand{\av}{$A_{\mathrm{V}}$}
\newcommand{\logg}{$\log{g}$}
\newcommand{\settl}{Settl-Net}
\newcommand{\nextgen}{NextGen-Net}
\newcommand{\dust}{Dusty-Net}

\newcommand{\hii}{\mbox{H\,\textsc{ii}}}
\newcommand{\nai}{\mbox{Na\,\textsc{i}}}
\newcommand{\ki}{\mbox{K\,\textsc{i}}}
\newcommand{\cai}{\mbox{Ca\,\textsc{i}}}
\newcommand{\caii}{\mbox{Ca\,\textsc{ii}}}
\newcommand{\mgi}{\mbox{Mg\,\textsc{i}}}
\newcommand{\baii}{\mbox{Ba\,\textsc{ii}}}
\newcommand{\fei}{\mbox{Fe\,\textsc{i}}}

\newcommand{\ha}{H$\mathrm{\alpha}$}   

\usepackage[colorlinks=true, citecolor=blue, linkcolor=blue]{hyperref}
\begin{document}

   \title{Spectral classification of young stars using conditional invertible neural networks}

   \subtitle{I. Introducing and validating the method}

   \author{Da Eun Kang\inst{1}
          \and
          Victor F. Ksoll\inst{1}
          \and
          Dominika Itrich\inst{3,4}
          \and
          Leonardo Testi\inst{5,6}
          \and
          Ralf S. Klessen\inst{1,2}
          \and
          Patrick Hennebelle\inst{7}
          \and
          Sergio Molinari\inst{8}
          }

   \institute{Universit\"{a}t Heidelberg, Zentrum f\"{u}r Astronomie, 
            Institut f\"{u}r Theoretische Astrophysik, 
            Albert-Ueberle-Stra{\ss}e 2,\\
            D-69120 Heidelberg, Germany\\
            \email{kang@uni-heidelberg.de}
            \and
            Universit\"{a}t Heidelberg, Interdisziplin\"{a}res Zentrum f\"{u}r Wissenschaftliches Rechnen, Im Neuenheimer Feld 205,\\
            D-69120 Heidelberg, Germany
            \and
            European Southern Observatory, Karl-Schwarzschild-Str. 2, 85748 Garching bei M\"{u}nchen, Germany
            \and
            Universit\"{a}ts-Sternwarte, Ludwig-Maximilians-Universit\"{a}t, Scheinerstrasse 1, 81679 M\"{u}nchen, Germany
            \and
            Alma Mater Studiorum Università di Bologna, Dipartimento di Fisica e Astronomia (DIFA), Via Gobetti 93/2, I-40129, \\
            Bologna, Italy
            \and
            INAF-Osservatorio Astrofisico di Arcetri, Largo E. Fermi 5, I-50125, Firenze, Italy 
            \and 
            Université Paris Cité, Université Paris-Saclay, CEA, CNRS, AIM, F-91191, Gif-sur-Yvette, France
            \and
            INAF - Istituto di Astrofisica e Planetologia Spaziali, Via Fosso del Cavaliere 100, I-00133 Roma, Italy
             }

   \date{Received XXX; accepted YYY}

 
  \abstract
  {}
   {We introduce a new deep learning tool that estimates stellar parameters (such as effective temperature, surface gravity, and extinction) of young low-mass stars by coupling the Phoenix stellar atmosphere model with a conditional invertible neural network (cINN). Our networks allow us to infer the posterior distribution of each stellar parameter from the optical spectrum. }
   {We discuss cINNs trained on three different Phoenix grids: Settl, NextGen, and  Dusty. We evaluate the performance of these cINNs on unlearned Phoenix synthetic spectra and on the spectra of 36 Class III template stars with well-characterised stellar parameters.
   }
   {We confirm that the cINNs estimate the considered stellar parameters almost perfectly when tested on unlearned Phoenix synthetic spectra. Applying our networks to Class III stars, we find good agreement with deviations of at most 5--10 per cent.
    The cINNs perform slightly better for earlier-type stars than for later-type stars like late M-type stars, but we conclude that estimations of effective temperature and surface gravity are reliable for all spectral types within the network's training range. 
    }
   {Our networks are time-efficient tools applicable to large amounts of observations. Among the three networks, we recommend using the cINN trained on the Settl library (Settl-Net), as it provides the best performance across the largest range of temperature and gravity.
   }


   \maketitle
%

\section{Introduction}
\label{sec:introduction}

In star-forming regions, it is massive stars that influence the surrounding environment energetically and dynamically during their short lifetime, but the majority of stars formed in star-forming regions are low-mass stars similar to or less than the solar mass. These low-mass stars are not only the most numerous in the star-forming region~\citep{Bochanski+10} but also account for about half of the total stellar mass~\citep{Kroupa02, Chabrier03}. Living longer than massive stars, these low-mass stars still remain in the pre-main-sequence phase even when the massive stars are dead. These young low-mass stars provide important information for studying stellar evolution and planet formation.

Stellar parameters (e.g. effective temperature, surface gravity, luminosity, etc.) are estimated from photometric or spectroscopic data by various methods. These methods are usually based on characteristic spectral features that vary depending on the type of stars. Therefore, it is important to adopt the appropriate method to the star under consideration and to the observed wavelength range.

As the volume of accumulated observations ever-expand in recent days, it has become important to develop time-efficient tools that analyse large amounts of data in a faster and more consistent way. This is why artificial neural networks~\citep[NNs;][]{Goodfellow+16} have been utilised in many astronomical fields these days. For instance, NNs have been used to predict physical parameters~\citep[e.g.][]{Fabbro+18, Ksoll+20,Olney2020, Kang+22} or to efficiently analyse images such as identifying structures~\citep[e.g.][]{Abraham+18} and exoplanets~\citep[e.g.][]{DeBeurs2022} or classifying observations~\citep[e.g.][]{Wu+19, Walmsley+21,Whitmore+21}.
In this study, we develop NNs that can efficiently analyse numerous spectra in the optical wavelength range of young low-mass stars. We prepare our networks to analyse VLT/MUSE observations adopting the wavelength coverage and spectral resolution of MUSE. In the follow-up paper, we will apply our tool to the spectra of young stars in the Carina Nebula observed with VLT/MUSE.

We adopt conditional invertible neural network (cINN) architecture developed by \cite{Ardizzone+21}. Estimating physical parameters from observed measurements is a non-trivial task. As the information we obtain from observation is limited due to the information loss during the forward process (i.e. translation from physical systems to observations), different physical systems can be observed similarly or almost identically, which we call a degenerate system. cINN architecture is specialised for solving the inverse problem of the degenerate system (i.e. from observations to physical systems). In particular, cINN has its own advantage in that cINN always provides a full posterior distribution of the physical system without any additional computations. In astronomy, the cINN approach has been used so far to characterise the internal properties of planets \citep{Haldemann2022}, analyse photometric data of young stars~\citep{Ksoll+20}, study emission lines in \hii\ regions~\citep{Kang+22}, or infer the merger history of galaxies \citep{Eisert2023}.

The cINN architecture adopts a supervised learning approach that learns the hidden rules from a number of well-labelled data sets of physical parameters and observations. As it is difficult to collect a sufficient number of well-interpreted real observations, synthetic observations have been usually used instead to generate enough training data. In this study, we utilise Phoenix stellar atmosphere libraries~\citep[e.g.][]{Allard2012,Husser2013,Baraffe2015} to train cINNs. Selecting Settl, NextGen, and Dusty Phoenix libraries, we introduce three cINNs (\settl, \nextgen, and \dust) trained on each library.

A few studies have developed NNs to analyse low-mass stars from photometric or spectroscopic data~\citep[e.g.][]{Ksoll+20, Olney2020, Sharma2020}. For example, \cite{Ksoll+20} developed a network using cINN architecture to estimate the physical parameters of individual stars from HST photometric data and \cite{Olney2020} used a convolutional neural network (CNN) to estimate physical parameters (e.g. effective temperature, surface gravity, and metallicity) from near-infrared spectra observed with Apache Point Observatory Galactic Evolution Experiment (APOGEE) spectrograph. \cite{Sharma2020} used CNN as well to diagnose the optical spectra of stars in a wide range of spectral types but their network only estimates the spectral type of the stars, not the other physical parameters. 
On the other hand, in this paper, our networks directly estimate the stellar parameters from the optical spectrum of low-mass stars, including the stars in both the main sequence and pre-main sequence phases. Moreover, our network provides a posterior distribution by adopting cINN architecture, which is useful to study the degeneracy between parameters.

In this paper, we focus on validating the performance of three cINNs. We evaluate our networks not only on Phoenix synthetic observations but also on real spectra of 36 young low-mass stars to investigate how well our cINNs work on real observations. These stars are template stars in the Class III phase, well-interpreted by literature~\citep[e.g.][]{Manara2013, Manara2017, Stelzer2013}.

The paper is structured as follows. In Sect.~\ref{sec:nn}, we describe the structure and principles of cINN and explain implementation detail on the machine learning side. In Sect.~\ref{sec:train}, we introduce our three networks and three training databases. In the following section (Sect.~\ref{sec:template}), we describe the Class III template stars used in this paper. Our main results are in Sect.~\ref{sec:validation}. We validate our networks using synthetic Phoenix spectra and 36 template stars. We not only evaluate the parameter prediction power of the cINN but also check whether the predicted parameters do explain the input observations. Section~\ref{sec:featureimportance} present which parts of the spectrum cINN relies mostly upon. In Sect.~\ref{sec:simulation_gap}, we investigate the gap between Phoenix synthetic spectra and real observations. We summarise the results in Sect.~\ref{sec:summary}.

\section{Neural network}
\label{sec:nn}
\subsection{Conditional invertible neural network}
\label{sec:cinn}

The conditional invertible neural network \citep[cINN;][]{Ardizzone2019a, Ardizzone2019b} is a deep learning architecture that is well suited for solving inverse problems that are tasks where the underlying physical properties $\mathbf{x}$ of a system are to be recovered from a set of observable quantities $\mathbf{y}$. In nature, recovering the inverse mapping $\mathbf{x} \leftarrow \mathbf{y}$ is often challenging and subject to degeneracy due to an inherent loss of information in the forward mapping $\mathbf{x} \rightarrow \mathbf{y}$, such that multiple sets of physical properties may appear similar or even entirely the same in observations. 

To tackle these difficulties, the cINN approach introduces a set of unobservable, latent variables $\mathbf{z}$ with a known, prescribed prior distribution $P(\mathbf{z})$ to the problem in order to encode the information that is otherwise lost in the forward mapping. The cINN achieves this by learning a mapping $f$ from the physical parameters $\mathbf{x}$ to the latent variables $\mathbf{z}$ \textit{conditioned} on the observations $\mathbf{y}$, that is
\begin{equation}
    f\left(\mathbf{x}; \mathbf{c} = \mathbf{y}\right) = \mathbf{z}, 
\end{equation}
capturing all the variance of $\mathbf{x}$ not explained by $\mathbf{y}$ in $\mathbf{z}$, while enforcing that $\mathbf{z}$ follows the prescribed prior $P(\mathbf{z})$. Given a new observation $\mathbf{y}'$ at prediction time, the cINN can then query the encoded variance by sampling the latent space according to the known prior distribution and by making use of its invertible architecture run in reverse to estimate the full posterior distribution $p\left(\mathbf{x}|\mathbf{y}'\right)$ as
\begin{equation}
    p\left(\mathbf{x}|\mathbf{y}'\right) \sim g\left(\mathbf{z}; c = \mathbf{y}'\right), \,\, \mathrm{with}\,\, \mathbf{z} \propto P\left(\mathbf{z}\right),
\end{equation}
where $f^{-1}(\cdot, \mathbf{c}) = g(\cdot, \mathbf{c})$ represents the inverse of the learned forward mapping for fixed condition $\mathbf{c}$. In practise $P(\mathbf{z})$ is usually prescribed to be a multivariate normal distribution with zero mean and unit covariance, and the dimension of the latent space is chosen to be equal to that of the target parameter space, that is~$\dim(\mathbf{z}) = \dim(\mathbf{x})$.

The invertibility of the cINN architecture is achieved by chaining so-called (conditional) affine coupling blocks \citep{Dinh2016}. Each of these blocks performs two complementary affine transformations on the halves $\mathbf{u}_1$ and $\mathbf{u}_2$ of the block input vector $\mathbf{u}$, following
\begin{equation}
    \label{eq:acb_forward}
    \begin{split}
        \mathbf{v}_1 &= \mathbf{u}_1 \odot \exp\left(s_2(\mathbf{u}_2, \mathbf{c})\right) \oplus t_2(\mathbf{u}_2, \mathbf{c}) \\
        \mathbf{v}_2 &= \mathbf{u}_2 \odot \exp\left(s_1(\mathbf{v}_1, \mathbf{c})\right) \oplus t_1(\mathbf{v}_1, \mathbf{c}).
    \end{split}
\end{equation}
As the equation shows, these two transformations are easily inverted given the halves $\mathbf{v}_1$, $\mathbf{v}_2$ of the output vector $\mathbf{v}$ according to
\begin{equation}
    \label{eq:acb_backward}
    \begin{split}
        \mathbf{u}_2 &= \left(\mathbf{v}_2 \ominus t_1(\mathbf{v}_1, \mathbf{c})\right) \odot \exp\left(-s_1(\mathbf{v}_1, \mathbf{c})\right) \\
        \mathbf{u}_1 &= \left(\mathbf{v}_1 \ominus t_2(\mathbf{u}_2, \mathbf{c})\right) \odot \exp\left(-s_2(\mathbf{u}_2, \mathbf{c})\right).
    \end{split}
\end{equation}
In both sets of Eqs. \eqref{eq:acb_forward} and \eqref{eq:acb_backward}, $s_i$ and $t_i$ ($i \in \{1, 2\}$) denote arbitrarily complex transformations, which need not themselves be invertible (as they are only ever evaluated in the forward direction) and can also be learned by the cINN itself when realised as small sub-networks \citep{Ardizzone2019a, Ardizzone2019b}.

Another advantage of the cINN architecture is that, as the observations are treated as a condition and simply concatenated to the input of the subnetworks $s_i$ and $t_i$ in each affine coupling layer, it allows for a) an arbitrarily large dimension of the input $\mathbf{y}$ and b) the introduction of a conditioning network $h$ (trained together with the cINN itself), which transforms the input observation into a more helpful, learned representation $\tilde{\mathbf{y}} = h(\mathbf{y})$ for the cINN \citep{Ardizzone2019b}. 

\subsection{Implementation details}
\label{sec:implementation}
In this paper, we employ a cINN consisting of 11--16 conditional affine coupling layers in the GLOW \citep[Generative Flow;][]{Kingma2018} configuration, where the transformation outputs $s_i(\cdot)$ and $t_i(\cdot)$ are estimated by a single subnetwork $r_i(\cdot) = (s_i(\cdot), t_i(\cdot))$. The latter choice, reduces the number of sub-networks per affine layer from four to two, reducing network complexity and computation time. As sub-networks $r_i$ we employ simple fully-connected architectures consisting of 5--7 layers of size $256$ using the rectified linear unit (ReLU, $\mathrm{ReLU}(x) = \max(0, x)$) as activation function. 

The affine coupling layers are, furthermore, alternated with random permutation layers, which randomly (but in a fixed and, thus, invertible way) permute the output vector in between coupling layers to improve the mixing of information between the two streams $\mathrm{u}_1$ and $\mathrm{u}_2$ \citep{Ardizzone2019a, Ardizzone2019b}. For the conditioning network $h$, we also employ a three-layer fully-connected architecture with layer size $512$ and ReLU activation, extracting $256$ features in the final layer.

Prior to training, we perform a linear scaling transformation on both the target parameters $\mathbf{x} = \{x_1, \ldots, x_N\} $ and input observations $\mathbf{y} = \{y_1, \ldots, y_M\}$, where each target property $x_i$ and input feature $y_i$ is modified according to 
\begin{equation}
    \begin{split}
        \hat{x}_i &= \frac{x_i - \mu_{x_i}}{\sigma_{x_i}}, \\
        \hat{y}_i &= \frac{y_i - \mu_{y_i}}{\sigma_{y_i}},
    \end{split}
\end{equation}
where $\mu_{x_i}$, $\mu_{y_i}$ and $\sigma_{x_i}$, $\sigma_{y_i}$, denote the means and standard deviations of the respective parameter/feature across the training data set. These transformations ensure that the distributions of individual target parameters/input features have zero mean and unit standard deviation, and are trivially inverted at prediction time. The transformation coefficients $\mu_{x_i}$, $\mu_{y_i}$ and $\sigma_{x_i}$, $\sigma_{y_i}$ are determined from the training set and applied in the same way to new query data. 

We train the cINN approach for this problem by minimisation of the maximum likelihood loss as described in \citep{Ardizzone2019b} using the Adam \citep{Kingma2014} optimiser for stochastic gradient descent with a step-wise learning rate adjustment.


\section{Training data}
\label{sec:train}

\subsection{Stellar photosphere models}
\label{sec:spectral_libraries}

The approach used to train the cINN is to use libraries of theoretical models for stellar photospheres. Our goal is to use the cINN to be able to classify and derive photospheric parameters from medium to low-resolution optical spectroscopy. To this purpose, we selected the most extensive set of available models that offer a spectral resolution better than R$\sim$10000.
The most extensive, homogeneous, tested, and readily available\footnote{We downloaded the theoretical spectra from the websites: \href{url}{https://osubdd.ens-lyon.fr/phoenix/} and \href{url}{http://svo2.cab.inta-csic.es/theory/newov2/}} library of theoretical photospheric spectra, including different treatments of dust and molecules formation and opacities, applicable in the range of effective temperatures covering the range from $\sim$2000 to $\sim$7000~K and gravities appropriate for pre-main sequence stars and brown dwarfs are the Phoenix spectral libraries \citep[e.g.][]{Allard2012, Husser2013, Baraffe2015}.
In this study, we have used the NextGen, Dusty, and Settl models, the latter is expected to provide the best description of the atmospheric characteristics in most cases of interest \citep{Allard2012}. We have included the older NextGen models as a comparison set, and the Dusty models as they seem to more accurately describe photospheres in the range of $2000~{\rm K}\le T_{\text{eff}}\le 3000$~K \citep[e.g.][]{Testi2009}. For a more detailed description and comparison of the physical assumption in the models, see the discussion and references in \citet{Allard2012}.

The grid of synthetic spectra is available for regularly spaced values of \teff\ and $\log{g}$, with steps of 100~K in \teff\ and 0.5 in $\log{g}$. To compute a synthetic spectrum for a given set of (arbitrary but within the grid ranges) values of (\teff, $\log{g}$, and \av) we set up the following procedure:
first, we identify the values of \teff\ and $\log{g}$ in the grid that bracket the requested values, then we interpolate linearly in $\log{g}$ at the values of the two bracketing \teff\ values, then we interpolate linearly the two resulting spectra at the requested \teff\ value, finally, we compute and apply the extinction following the \citet{Cardelli1989} prescription, with $R_{\mathrm{V}}$ as a user selectable parameter (in this study we use $R_{\mathrm{V}}$=4.4, see Sect.~\ref{sec:training_data}).
The resulting spectrum is then convolved at the MUSE resolution, using a Gaussian kernel, and resampled on the MUSE wavelength grid.

\subsection{Databases and networks}
\label{sec:training_data}

In this study, we analyse the cINN performance based on each of the three spectral libraries described in the previous section. Accordingly, we construct a training data set for each spectral library using the interpolation scheme we have outlined. For the target parameter space, we adopt the following limits: 

For NextGen and Settl we limit $T_\mathrm{eff}$ to a range of $2600$ to $7000\,$K and $\log(g/\mathrm{cm\,s}^{-2})$ from $2.5$ to $5$. The Dusty library has an overall smaller scope, therefore we can only probe from $2600$ to $4000\,$K in $T_\mathrm{eff}$ and from $3$ to $5$ in $\log(g/\mathrm{cm\,s}^{-2})$ here. For $A_{\mathrm{V}}$ we select the same range of $0$ to $10$ mag for all three libraries, where we use the \cite{Cardelli1989} extinction law with $R_{\mathrm{V}} = 4.4$ to artificially redden the model spectra. We choose $R_{\mathrm{V}} = 4.4$ considering the application of our networks to the Carina Nebula~\citep{Hur+2012} in the follow-up study. As some of the template stars used in this paper (Sect.~\ref{sec:template}) are dereddend assuming $R_{\mathrm{V}}=3.1$, we have also experimented with training data sets using $R_{\mathrm{V}}=3.1$. We have not found a significant difference in our main results, therefore we keep using $R_{\mathrm{V}} = 4.4$ in this study.

In terms of wavelength coverage, we match the range of the template spectra described in Sect.~\ref{sec:template}, (i.e. $\sim5687$ to $\sim9350\,$\AA) and adopt the MUSE spectral resolution subdividing the wavelength interval into a total of 2930 bins with a width of $1.25\,$\AA. Additionally, we normalise the spectra to the sum of the total flux across all bins. 

To generate the training data we opt for a uniform random sampling approach, where we sample both $T_\mathrm{eff}$ and $g$ in log space and only $A_{\mathrm{V}}$ in linear space within the above-specified limits for the three libraries. We generate a total of 65,536 synthetic spectra models for each library. Note that we have also experimented with larger training sets, but have not found a significant increase in the predictive performance of our method, such that we deemed this training set size sufficient. 

Finally, we randomly split each of these three initial databases 80:20 into the respective training and test sets for the cINN. The former subsets mark the data that the cINN is actually trained on, whereas the latter are with-held during training and serve to quantify the performance of the trained cINN on previously unseen data with a known ground truth of the target parameters.

We first train 50 networks for each library with randomised hyper-parameters of cINN and we select the best network based on the performance on the test set and template stars. We train the network until both training loss and test loss converge or either of them diverges, where the latter cases are discarded. It takes about 50 min to train one network (6 hours for 50 networks using 7 processes in parallel) with an NVIDIA GeForce RTX 2080 Ti graphic card. Once trained, our networks can sample posterior estimates very efficiently. Using the same graphic card (NVIDIA GeForce RTX 2080 Ti graphic card) and sampling 4096 posterior estimates per observation, it takes about 1.1 sec to sample posterior distributions for 100 observations (91 observations per second). When tested with M1 pro CPU with 8 cores, it takes about  13 sec for 100 observations (7.6 observation/sec).


\section{Class III templates}
\label{sec:template}

The set of observations on which we validate our networks contains 36 spectra of well-known Class III stars observed with VLT/X-Shooter \citep{Manara2013, Manara2017}. We refer the reader for details of observations and data reduction to original papers. 
Templates come from different star-forming regions (Taurus, Lupus, Upper Scorpius, $\sigma$ Orionis, TW Hydrae Association, Chameleon I) 
and span a broad range of effective temperatures (2300 -- 5800~K), as well as spectral types (M9.5 - G5.0). In this work we use their properties provided by \cite{Manara2013, Manara2017, Stelzer2013}. 

Spectral types for stars later than K5 were obtained based on the depth of molecular absorption bands (TiO, VO and CaH) and a few photospheric lines (e.g. \nai, \cai, \mgi, etc.) present in the optical part of the spectra \citep{Manara2013}. Earlier K-type stars were identified using the spectral indices introduced by \cite{Herczeg2014}, while G-type stars were identified based on the difference at 5150~\AA\ of continuum estimated between 4600 and 5400~\AA, and 4900 and 5150~\AA~\citep{Herczeg2014}. Effective temperatures (\teff) were derived from spectral types using relations from \cite{Luhman2003} for M-type objects and \cite{Kenyon1995} for K- and G-type stars. Most of the templates have none or negligible extinction \citep[$A_{\mathrm{V}}<0.5$~mag,][]{Manara2017}; those with $A_{\mathrm{V}}>0.3$ were dereddened before analysis assuming the extinction law from \cite{Cardelli1989} and $R_{\mathrm{V}}=3.1$. 

Surface gravity ($\log{g}$) of Class III sources was estimated using the ROTFIT tool \citep{Frasca2003}. It compares the observed spectrum with the grid of referenced spectra and finds a best-fit minimising the $\chi^2$ of difference between the spectra in specific wavelength ranges. \cite{Stelzer2013} and \cite{Manara2017} used BT-Settl spectra in a $\log{g}$ range of 0.5 -- 5.5 dex as reference. The tool also provides \teff, radial and rotational velocities; but we use \teff\ derived from spectral types in the subsequent analysis. Table~\ref{tab:class_III_templates} provides a summary of the Class III stars and their stellar parameters. 
We exclude from the original paper sources, which are suspected to be unresolved binaries or their youth is doubtful due to the lack of lithium absorption line at 6708~\AA~\citep{Manara2013}.

X-Shooter has higher spectral resolution than MUSE, thus template spectra were degraded to MUSE resolution (R$\sim$4000) using a Gaussian kernel and re-sample on MUSE spectra within the range of 5687.66 -- 9348.91~\AA~(the common spectral range of MUSE and optical arm of X-Shooter). Subsequently, spectra are normalised to the sum of the total flux of the stellar spectrum within the analysed spectral range.

\begin{table*}
    \centering
    \caption{Stellar parameters of Class III template stars.}
    \begin{tabular}{llcccc}
    \toprule
                    Object Name &  Region &  Spectral Type &  $T_\mathrm{eff} (K)$ &  $\log(g/\mathrm{cm\,s}^{-2})$ & Reference $\log(g)$  \\
    \midrule
          RXJ0445.8+1556 &     Taurus & G5.0 &  5770 &  3.93 &        (1) \\
          RXJ1508.6-4423 &     Lupus & G8.0 &  5520 &  4.06 &        (1) \\
          RXJ1526.0-4501 &     Lupus & G9.0 &  5410 &  4.38 &        (1) \\
                  HBC407 &     Taurus & K0.0 &  5110 &  4.33 &        (1) \\
    PZ99J160843.4-260216 &    Upper Scorpius & K0.5 &  5050 &  3.48 &        (1) \\
          RXJ1515.8-3331 &     Lupus & K0.5 &  5050 &  3.86 &        (1) \\
    PZ99J160550.5-253313 &    Upper Scorpius & K1.0 &  5000 &  3.81 &        (1) \\
          RXJ0457.5+2014 &     Taurus & K1.0 &  5000 &  4.51 &        (1) \\
          RXJ0438.6+1546 &     Taurus & K2.0 &  4900 &  4.12 &        (1) \\
          RXJ1547.7-4018 &     Lupus & K3.0 &  4730 &  4.22 &        (1) \\
          RXJ1538.6-3916 &     Lupus & K4.0 &  4590 &  4.21 &        (1) \\
          RXJ1540.7-3756 &     Lupus & K6.0 &  4205 &  4.42 &        (1) \\
          RXJ1543.1-3920 &     Lupus & K6.0 &  4205 &  4.12 &        (1) \\
                   SO879 & $\sigma$ Orionis & K7.0 &  4060 &  3.90 &       (2) \\
            Tyc7760283\_1 &   TW Hydrae & M0.0 &  3850 &   4.70 &         (2) \\
                   TWA14 &   TW Hydrae & M0.5 &  3780 &  4.70 &       (2) \\
     RXJ1121.3-3447\_app2 &   TW Hydrae & M1.0 &  3705 &   4.60 &         (2) \\
     RXJ1121.3-3447\_app1 &   TW Hydrae & M1.0 &  3705 &   4.80 &         (2) \\
             CD\_29\_8887A &   TW Hydrae & M2.0 &  3560 &   4.40 &         (2) \\
             CD\_36\_7429B &   TW Hydrae & M3.0 &  3415 &   4.50 &         (2) \\
              TWA15\_app2 &   TW Hydrae & M3.0 &  3415 &  4.60 &       (2) \\
                    TWA7 &   TW Hydrae & M3.0 &  3415 &  4.40 &       (2) \\
              TWA15\_app1 &   TW Hydrae & M3.5 &  3340 &  4.50 &       (2) \\
                   SO797 & $\sigma$ Orionis & M4.5 &  3200 &  3.90 &       (2) \\
                   SO641 & $\sigma$ Orionis & M5.0 &  3125 &  3.80 &       (2) \\
              Par\_Lup3\_2 &     Lupus & M5.0 &  3125 &  3.70 &       (2) \\
                   SO925 & $\sigma$ Orionis & M5.5 &  3060 &  3.80 &       (2) \\
                   SO999 & $\sigma$ Orionis & M5.5 &  3060 &  3.80 &       (2) \\
                   Sz107 &     Lupus & M5.5 &  3060 &  3.70 &       (2) \\
              Par\_Lup3\_1 &     Lupus & M6.5 &  2935 &  3.60 &       (2) \\
                   LM717 &    Chameleon I & M6.5 &  2935 &  3.50 &       (2) \\
       J11195652-7504529 &    Chameleon I & M7.0 &  2880 &  3.09 &        (1) \\
                   LM601 &    Chameleon I & M7.5 &  2795 &   4.00 &         fixed \\
               CHSM17173 &    Chameleon I & M8.0 &  2710 &   4.00 &         fixed \\
                   TWA26 &   TW Hydrae & M9.0 &  2400 &  3.60 &       (2) \\
               DENIS1245 &   TW Hydrae & M9.5 &  2330 &   3.60 &       (2) \\
    \bottomrule
    \end{tabular}
    \label{tab:class_III_templates} 
    \tablefoot{The last column indicates the literature source of the $\log(g)$ values, where "fixed" indicates that no measurement was available in the literature and we assumed a fixed value of $\log(g/\mathrm{cm\,s}^{-2}) = 4.0$ instead.}
    \tablebib{(1)~\citet{Manara2017}; (2) \citet{Stelzer2013}.}
\end{table*}

\section{Validation}
\label{sec:validation}

\subsection{Validations with synthetic spectra}
\label{sec:validation_synthetic}

\renewcommand{\arraystretch}{1.25}
\begin{table*}
    \centering
    \caption{ Average prediction performance of three networks (\settl, \nextgen, and \dust) on 13,107 Phoenix synthetic models in the test set.
    \label{table:nrmse_3nets}}
    \begin{tabular}{ l  c  c  c   c c c}
        \toprule
        \multicolumn{1}{l}{} & \multicolumn{3}{c}{RMSE} & \multicolumn{3}{c}{NRMSE} \\
        \cmidrule(rl){2-4} \cmidrule(rl){5-7}
        Network & log $T_{\mathrm{eff}}$ & $\log(g)$ & $A_{\mathrm{V}}$ &
                log $T_{\mathrm{eff}}$ & $\log(g)$ & $A_{\mathrm{V}}$ \\
        \midrule
        Settl  &  $4.260 \times 10^{-4}$  &  $1.211 \times 10^{-2}$  &  $7.893 \times 10^{-3}$  &  $9.904 \times 10^{-4}$  &  $4.846 \times 10^{-3}$  &  $7.893 \times 10^{-4}$   \\ 
        NextGen  &  $3.064 \times 10^{-4}$  &  $6.742 \times 10^{-3}$  &  $6.499 \times 10^{-3}$  &  $7.123 \times 10^{-4}$  &  $2.697 \times 10^{-3}$  &  $6.499 \times 10^{-4}$   \\ 
        Dusty  &  $7.274 \times 10^{-5}$  &  $1.573 \times 10^{-3}$  &  $2.517 \times 10^{-3}$  &  $3.888 \times 10^{-4}$  &  $7.863 \times 10^{-4}$  &  $2.517 \times 10^{-4}$   \\ 

        \bottomrule
    \end{tabular} 
    \tablefoot{For each parameter and each network, we present the RMSE, the mean accuracy of the MAP estimates, and the RMSE normalised by the parameter range covered in the training data (NRMSE). The test set of each network is drawn from the corresponding synthetic database.}
\end{table*}

\begin{figure*}
	\includegraphics[width=\textwidth]{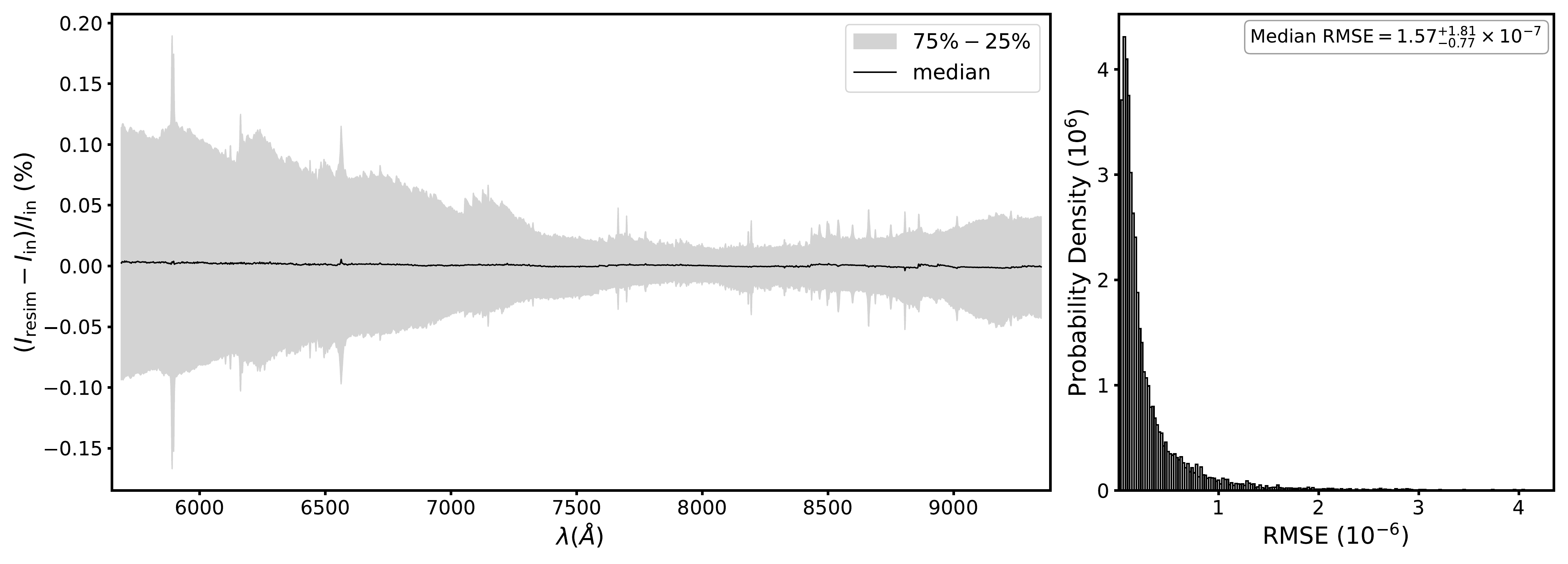}
    \caption{Resimulation results of \settl\ for the entire synthetic spectra in the test set.
    The left panel presents the median relative error across the wavelength range of the resimulated spectra based on the MAP predictions of the cINN trained on the Settl models averaged over the 13,107 synthetic spectra in the test set. Here the grey envelope indicates the interquantile range between the $25\%$ and $75\%$ quantiles. In the right panel, we present the histogram of the RMSEs of the 13,107 resimulated spectra. The mean resimulation RMSE across the test set is $3.01 \pm 4.35 \times 10^{-7}$. }
    \label{fig:ResimSynth_Settl}
\end{figure*}

In this section, we validate whether the trained networks well learned the physical rules hidden in the synthetic Phoenix models or not. We use the test set of each database, the synthetic models that are not used for the training but share the same physics as the training data. As mentioned in Sect. \ref{sec:training_data}, we only used 80\%\ of the database for training and remained the rest for validation. Each test set consists of 13,107 test models.

\subsubsection{Prediction performance}
\label{sec:predictionSynth}
We introduce an accuracy index for evaluating the parameter prediction performance of the network. The accuracy of the prediction is defined as the deviation between the posterior estimate of the parameter and the ground true value ($x^{*}$) of the test model. 
In this section, we calculate the accuracy in the same physical scales we used to build the databases in Sect.~\ref{sec:training_data}, meaning that we use the logarithmic scales for the effective temperature and surface gravity and use the linear scale for the extinction magnitude.
We use either all posterior estimates sampled for one test model or the maximum a posteriori (MAP) point estimate as a representative. To determine the MAP estimate from the posterior distribution, we perform a Gaussian kernel density estimation on a 1D posterior distribution and find the point where the probability density maximises, similar to the method used in \cite{Ksoll+20} and \cite{Kang+22}. In most parts of this paper, we use the MAP estimate to quantify the accuracy of the prediction.

We evaluate the three networks (\settl, \nextgen, and \dust) by using all 13,107 test models in the corresponding test set. For each test model, we sample 4096 posterior estimates and measure the MAP estimates for three parameters from the 1D posterior distributions.
In Fig.~\ref{fig:settl_test_map}, we present 2D histograms comparing the MAP values estimated by the \settl\ with the true values of the entire test models. The \settl\ predicts all three parameters extremely well so the data points are all lying on the 1-to-1 correspondence line. The \nextgen\ and \dust\ as well show extremely good results on the test set. The results of the other two networks are very similar to the result of \settl\ (Fig.~\ref{fig:settl_test_map}), so we do not include figures of them in this paper.

To quantify the average accuracy of the network for multiple test models, we measure the root mean square error (RMSE) following,
\begin{equation}
    \mathrm{RMSE} = \sqrt{\frac{\Sigma_{i=1}^{N}(x^{\mathrm{MAP}}_{i} - x^{*}_{i})^{2}    }{N}  }.
\end{equation}
In the case of the \dust, the training ranges of the effective temperature and surface gravity are narrower than that of the other two networks. As the total number of models is the same for all three databases (i.e. 65,536 models), the number density of the model for the effective temperature and surface gravity in the Dusty database is higher than the other two. On this account, we define the normalised RMSE (NRMSE),
\begin{equation}
    \mathrm{NRMSE} = \frac{ \mathrm{RMSE} }{ x_{\mathrm{max}}^{\mathrm{training}} -  x_{\mathrm{min}}^{\mathrm{training}}  },
\end{equation}
by dividing the RMSE by the training range.

In Table~\ref{table:nrmse_3nets}, we list the RMSE and NRMSE of each parameter for three networks. As already shown in the comparisons between the MAP values and true values (Fig.~\ref{fig:settl_test_map}), the RMSE and NRMSE for all three networks are very small around $10^{-4} \sim 10^{-2}$.
The \dust\ has the smallest RMSE and NRMSE for all three parameters among the three networks. In the case of the effective temperature and extinction, the differences in NRMSE between the networks are very small, whereas the difference in the NRMSE in the case of surface gravity is relatively noticeable among the three parameters.
Although the \dust\ has the best results, small values in Table~\ref{table:nrmse_3nets} demonstrate that all three networks perfectly learned about the synthetic spectra.

\subsubsection{Resimulation}
\label{sec:ResimSynth}

To further validate the prediction results of the cINN on the synthetic test data, we verify if the spectra that correspond to the MAP estimates match the respective input spectrum of each test example. We do so by feeding the MAP predictions for the stellar parameters of the 13,107 test examples as an input to our spectra interpolation routine, which we introduced for the training set generation in Sect.~\ref{sec:spectral_libraries}, in order to resimulate the corresponding spectra. Afterwards, we compute the residuals, RMSEs and $R^2$ scores of the resimulated spectra in comparison to the corresponding input spectra. The latter serves as a goodness-of-fit measure and is defined as
\begin{equation}
    R^2 = 1 - \frac{\sum_i (y_i - \hat{y}_i)^2}{\sum_i (y_i - \bar{y})^2}
\end{equation}
for a set of $N$ observations $y_i$ with corresponding predictions $\hat{y}_i$, where $\bar{y} = \frac{1}{N}\sum_i^N y_i$ denotes the mean of the observations. It takes on values between 0 and 1, with the latter indicating a perfect match \citep{James2017}. 

Figure~\ref{fig:ResimSynth_Settl} summarises the results for \settl, showing the median relative residual against the wavelength in the left panel and the distribution of RMSEs in the right one. The corresponding plots for \nextgen\ and \dust\ can be found in Figs.~\ref{fig:ResimSynth_NextGen} and \ref{fig:ResimSynth_Dusty} in the Appendix. Out of the 13,107 test cases, we could not resimulate spectra for only 52, 32 and 9 MAP predictions for the \settl, \nextgen\ and \dust, respectively. Only in these few instances either the predicted temperature or gravity (or both) fall outside the interpolation limits of the respective spectra library, such that the spectrum cannot be resimulated. Notably, all of these cases are extreme edge cases, that are right at the training boundaries of either $T_\mathrm{eff}$ or $\log(g)$ such that the cINN MAP estimates fall ever so slightly outside the limits while still being an excellent match to the ground truth. 

Figure~\ref{fig:ResimSynth_Settl} confirms the excellent precision of the MAP predictions demonstrated in the ground truth comparison in Fig.~\ref{fig:settl_test_map}. With a median RMSE of the resimulated spectra of $1.57_{-0.77}^{+1.81} \times 10^{-7}$ (and median $R^2$ score of $1$), we find that the resimulated spectra are practically spot-on to the corresponding input. In the left panel of Fig.~\ref{fig:ResimSynth_Settl} we can also see that, while the overall median residual is very small, there is a systematic trend towards a larger discrepancy between resimulation and input within a shorter wavelength regime ($<7250$ \AA). This is likely an effect of the overall low flux in the short wavelength regime for the colder stars ($< 4000$ K), such that even a small deviation in flux results in a comparably larger value of the relative residual. Although, it has to be noted again that with most relative deviations falling below $0.2\%$ the discrepancy is overall marginal even in the short wavelength regime. 

As Figs.~\ref{fig:ResimSynth_NextGen} and \ref{fig:ResimSynth_Dusty} show, \nextgen\ and \dust\ exhibit similar behaviour in the resimulation test, although we find slightly lower mean RMSEs with $2.28 \pm 2.48 \times 10^{-7}$ and $9.01 \pm 7.34 \times 10^{-8}$, respectively. Given that the mean RMSEs across the three different spectral libraries agree within one $\sigma$, however, it is safe to say that all three networks achieve equally excellent performance in the resimulation test.


\begin{table*}
    \centering
    \caption{Summary of cINN MAP predictions for the Class III template spectra for the cINN models based on the three different spectral libraries.}
    \resizebox{\textwidth}{!}{
    \begin{tabular}{lcccccccccc}
    \toprule
        & \multicolumn{9}{c}{MAP estimate} \\
        \cmidrule(rl){2-10} 
         & \multicolumn{3}{c}{$T_\mathrm{eff}$ (K) [$\Delta_\mathrm{lit}$]} & \multicolumn{3}{c}{$\log(g/\mathrm{cm\,s}^{-2})$ [$\Delta_\mathrm{lit}$]} &  \multicolumn{3}{c}{$A_{\mathrm{V}}$ (mag)} \\
        \cmidrule(rl){2-4} \cmidrule(rl){5-7} \cmidrule(rl){8-10}   
        Object Name & Settl & NextGen & Dusty & Settl & NextGen & Dusty & Settl & NextGen & Dusty \\
    \midrule
        RXJ0445.8+1556 & 5391 [379] & 5692 [78] & 4161 [1609] & 4.28 [-0.35] & 4.13 [-0.20] & 4.14 [-0.21] & 0.21 & 0.38 & -0.02 \\
        RXJ1508.6-4423 & 5069 [451] & 5434 [86] & 4141 [1379] & 4.10 [-0.04] & 4.16 [-0.10] & 4.13 [-0.07] & -0.31 & -0.04 & -0.13 \\
        RXJ1526.0-4501 & 5150 [260] & 5443 [-33] & 4170 [1240] & 4.25 [0.13] & 4.21 [0.17] & 4.13 [0.25] & -0.02 & 0.19 & 0.14 \\
        HBC407 & 5129 [-19] & 5497 [-387] & 4165 [945] & 4.71 [-0.38] & 4.64 [-0.31] & 4.26 [0.07]  & 0.17 & 0.37 & 0.02 \\
        PZ99J160843.4-260216 & 5006 [44] & 5366 [-316] & 4154 [896] & 4.43 [-0.95] & 4.42 [-0.94] & 4.28 [-0.80] & 0.15 & 0.38 & -0.09 \\
        RXJ1515.8-3331 &  4895  [155] & 5248 [-198] & 4177 [873] & 4.25 [-0.39] & 4.32 [-0.46] & 4.31 [-0.45] & 0.00 & 0.32 & 0.27 \\
        PZ99J160550.5-253313 &  4759  [241] & 5168 [-168] & 4192 [808] & 4.02 [-0.21] & 4.19 [-0.38] & 4.34 [-0.53] & 0.09 & 0.40 & 0.21 \\
        RXJ0457.5+2014 &  4644  [356] & 5105 [-105] & 4123 [877] & 4.37 [0.14] & 4.63 [-0.12] & 4.47 [0.04] & -0.13 & 0.34 & -0.17 \\
        RXJ0438.6+1546 &  4588  [312] & 4992 [-92] & 4177 [723] & 4.01 [0.11] &  4.20 [-0.08] & 4.50 [-0.38] & 0.01 & 0.44 & 0.20 \\
        RXJ1547.7-4018 &  4615  [115] & 5015 [-285] & 4185 [545] & 4.15 [0.07] & 4.40 [-0.18] & 4.52 [-0.30] & -0.02 & 0.26 & 0.13 \\
        RXJ1538.6-3916 &  4464  [126] & 4830 [-240] & 4180 [410] & 4.17 [0.04] & 4.38 [-0.17] & 4.69 [-0.48] & 0.01 & 0.30 & 0.21 \\
        RXJ1540.7-3756 &  4225  [-20] & 4260 [-55] & 4115 [90] & 4.22 [0.20] & 4.17 [0.25] & 4.92 [-0.50] & -0.11 & 0.12 & 0.22 \\
        RXJ1543.1-3920 &  4269  [-64] & 4299 [-94] & 4132 [73] & 4.34 [-0.22] & 4.32 [-0.20] & 5.00 [-0.88] & 0.03 & 0.28 & 0.39 \\
        SO879 &  4106  [-46] & 4027 [33] & 3909 [151] & 3.96 [-0.06] & 4.09 [-0.19] & 4.78 [-0.88] & 0.22 & 0.29 & -0.12 \\
        Tyc7760283\_1 &  3881  [-31] & 3748 [102] & 3742 [108] & 5.00 [-0.30] & 4.99 [-0.29] & 5.23 [-0.53] & -0.17 & -0.34 & -0.52 \\
        TWA14 &  3819  [-39] & 3739 [41] & 3677 [103] & 5.07 [-0.37] & 4.87 [-0.17] & 5.09 [-0.39] & -0.32 & 0.19 & -0.30 \\
        RXJ1121.3-3447\_app2 &  3797  [-92] & 3622 [83] & 3635 [70] & 4.78 [-0.18] & 4.68 [-0.08] & 5.13 [-0.53] & 0.38 & 0.30 & 0.02 \\
        RXJ1121.3-3447\_app1 &  3719  [-14] & 3559 [146] & 3564 [141] & 4.90 [-0.10] & 4.77 [0.03] & 5.16 [-0.36] & 0.01 & 0.04 & -0.07 \\
        CD\_29\_8887A &  3670 [-110] & 3483 [77] & 3491 [69] & 4.79 [-0.39] & 4.57 [-0.17] & 5.05 [-0.65] & 0.56 & 0.51 & 0.07 \\
        CD\_36\_7429B &  3423 [-8] & 3264 [151] & 3262 [153] & 4.70 [-0.20] & 4.44 [0.06] & 4.82 [-0.32] & 0.52 & 0.50 & 0.13 \\
        TWA15\_app2 &  3467 [-52] & 3289 [126] & 3306 [109] & 4.93 [-0.53] & 4.71 [-0.31] & 5.02 [-0.62] & 0.17 & 0.31 & 0.09 \\
        TWA7 &  3519 [-104] & 3321 [94] & 3316 [99] & 4.83 [-0.23] & 4.45 [0.15] & 4.80 [-0.20] & 0.41 & 0.94 & 0.14 \\
        TWA15\_app1 &  3469 [-129] & 3285 [55] & 3310 [30] & 5.01 [-0.51] & 4.79 [-0.29] & 5.08 [-0.58] & 0.06 & 0.20 & 0.10 \\
        SO797 &  3248 [-48] & 3225 [-25] & 3078 [122] & 3.93 [-0.03] & 3.47 [0.43] & 4.03 [-0.13] & 1.07 & 1.48 & 0.73 \\
        SO641 &  3129 [-4] & 3237 [-112] & 2997 [128] & 3.86 [-0.06] & 3.20 [0.60] & 3.81 [-0.01] & 0.68 & 1.46 & 0.43 \\
        Par\_Lup3\_2 &  3181 [-56] & 3245 [-120] & 3048 [77] & 3.96 [-0.26] & 3.29 [0.41] & 4.00 [-0.30] & 0.72 & 1.29 & 0.40 \\
        SO925 &  3008 [52] & 3277 [-217] & 2961 [99] & 3.76 [-0.06] & 2.92 [0.78] & 3.61 [0.09] & 0.97 & 2.01 & 0.76 \\
        SO999 &  3079 [-19] & 3294 [-234] & 2979 [81] & 3.68 [0.12] & 2.85 [0.95] & 3.58 [0.22] & 0.69 & 1.60 & 0.54 \\
        Sz107 &  2981 [79] & 3272 [-212] & 2935 [125] & 3.69 [0.11] & 2.85 [0.95] & 3.50 [0.30] & 0.56 & 1.67 & 0.35 \\
        Par\_Lup3\_1 &  2739 [196] & 3170 [-235] & 2868 [67] & 3.53 [-0.03] & 2.37 [1.13] & 3.04 [0.46] & 2.74 & 3.62 & 2.47 \\
        LM717 &  2714 [221] & 3218 [-283] & 2903 [32] & 3.46 [0.14] & 2.37 [1.23] & 2.84 [0.76] & 1.83 & 3.08 & 1.82 \\
        J11195652-7504529 &  2629 [251] & 3165 [-285] & 2864 [16] & 3.50 [-0.41] & 2.27 [0.82] & 2.75 [0.34] & 2.11 & 3.43 & 2.24 \\
        LM601 &  2601 [194] & 3137 [-342] & 2807 [-12] & 3.62 [-] & 2.28 [-] & 2.98 [-] & 1.79 & 3.16 & 2.00 \\
        CHSM17173 &  2539 [171] & 3096 [-386] & 2773 [-63] & 3.50 [-] & 2.18 [-] & 2.61 [-] & 1.66 & 3.45 & 2.31 \\
        TWA26 &  2477 [-77] & 2959 [-559] & 2625 [-225] & 3.46 [0.14] & 1.83 [1.77] & 2.56 [1.04] & 2.64 & 3.92 & 2.92 \\
        DENIS1245 &  2453 [-123] & 2924 [-594] & 2590 [-260] & 3.45 [0.15] &  1.71 [1.89] & 2.58 [1.02] & 2.34 & 3.74 & 2.82 \\
    \bottomrule
    \end{tabular}}
    \tablefoot{For $T_\mathrm{eff}$ and $\log(g)$ the value in parenthesis indicates the difference $x_\mathrm{lit} - x_\mathrm{MAP}$ to the literature stellar parameters listed in Table~\ref{tab:class_III_templates}. Since all Class III templates are assumed to be at zero extinction, for $A_{\mathrm{V}}$ the value itself is identical to the difference.}
    \label{tab:SummaryTemplateMAP}
\end{table*}

\subsection{Validations with Class III template stars}
\label{sec:validation_template}

In this section, we investigate how well our cINNs predict each parameter when applied to real observations by analysing Class III template stars introduced in Sect.~\ref{sec:template}. Stellar parameter values (i.e. effective temperature, surface gravity, and extinction) provided by previous papers~\citep{Manara2013, Manara2017,  Stelzer2013} are listed in Table~\ref{tab:class_III_templates}. Among the 36 template stars, there are cases where the literature value of effective temperature is out of the training range of the cINNs, or where the literature value of gravity is missing. Two out of 36 stars have temperatures below 2600~K, outside the temperature range of all three databases. Also, 14 stars with temperatures between 4000~K and 7000~K are out of the training range of the \dust. These stars will be excluded from some analyses in the following sections.

Using each network, we sample 4096 posterior estimates per star and measure MAP estimation for three parameters. We list the MAP values predicted by three networks in Table~\ref{tab:SummaryTemplateMAP}.

\begin{figure*}
	\includegraphics[width=2\columnwidth]{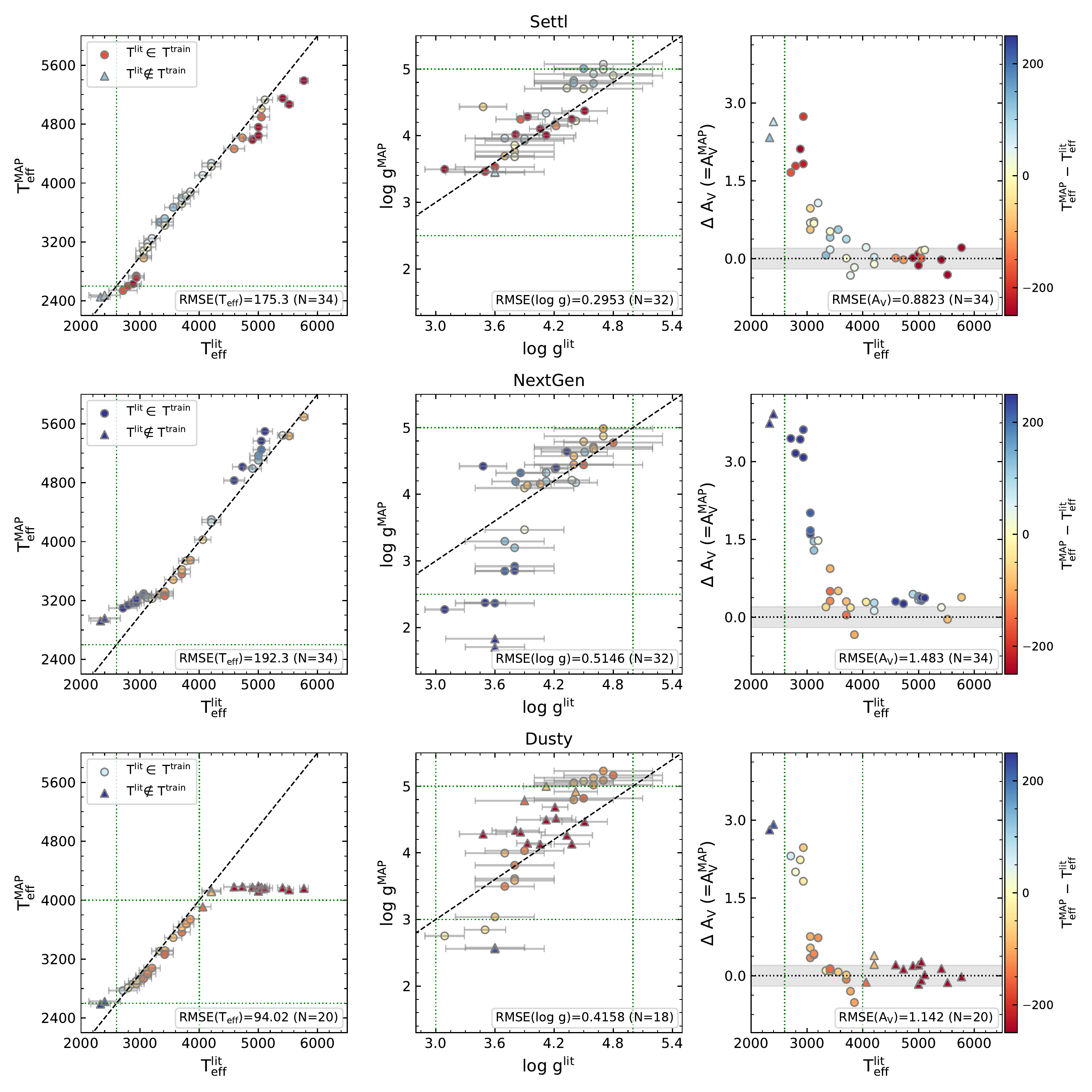}
    \caption{Comparison of MAP predictions with literature values in Table~\ref{tab:class_III_templates}. 
    Stars are basically denoted by circle symbols but triangle symbols denote stars excluded in analyses such as RMSE calculation either because their literature values of temperature are out of the cINN training range or because their literature values of surface gravity are missing. The colour indicates the temperature deviation between the MAP estimate and the literature value. 
    We indicate the training range of each parameter with green dotted lines.
    In the third column, the grey horizontal area presents the 1-$\sigma$ uncertainty (i.e. 0.2~mag) of extinction provided by literature.
     } \label{fig:tpl_cINN_pred}
\end{figure*}

\subsubsection{Parameter comparison between literature and cINN}
\label{sec:predictionTpl}

In Fig.~\ref{fig:tpl_cINN_pred}, we compare the stellar parameter values from literature ($x_\mathrm{lit}$) with MAP predictions ($x_\mathrm{MAP}$). Each row shows the result of different cINNs. The first two columns are the results of effective temperature and surface gravity. As the extinction value of template stars is negligible, we compare the literature value of the temperature with the MAP estimation of extinction.
We calculate the uncertainty of the MAP estimate based on the width of the posterior distribution, but as the uncertainties are all very small, we did not present the uncertainty of the MAP estimate in the figure.
For the uncertainty of the literature values, we adopt a 1-subclass temperature interval as the uncertainty of temperature and use the surface gravity uncertainty provided by the literature~\citep{Stelzer2013, Manara2017}. According to the literature, the 1-$\sigma$ uncertainty of extinction is $\sim$ 0.1--0.2~mag, so we indicate from $-0.2$ to $0.2$~mag range in grey to show the uncertainty range.

In this section, we do not use some stars in our analyses where the stellar parameter value from literature is out of the training range or where any stellar parameter value is missing, although they are presented in Fig.~\ref{fig:tpl_cINN_pred} by triangle symbol. We use 34, 34, and 20 stars for \settl, \nextgen, and \dust, respectively when analysing temperatures or extinction, and use 32, 32, and 18 stars respectively when analysing gravity.

Comparing the temperature MAP estimates with the literature values, we confirm that the majority of stars are lying close to the 1-to-1 correspondence line. We calculate the RMSE for each network by only using stars whose temperature literature values are within the training range (i.e. circle markers in Fig.~\ref{fig:tpl_cINN_pred}). 
Considering that the average of the 1-subclass temperature interval of these stars is about 140~K, the RMSE values of 175.3~K, 192.3~K, and 94.02~K for \settl, \nextgen, and \dust, respectively, are well within 1 to 2 subclasses interval.
As shown in the figure and RMSE values, \dust\ has the best agreement with the literature value when the temperature is within its training range of 2600 -- 4000~K. However, \dust\ shows very poor agreement with the literature values when the temperature is outside the training range. This implies the caution of using cINN to analyse stars far from the training range. Comparing \settl\ and \nextgen\ having the same training range, MAP estimates of \settl\ are closer to the literature values.

\begin{figure}
	\includegraphics[width=1\columnwidth]{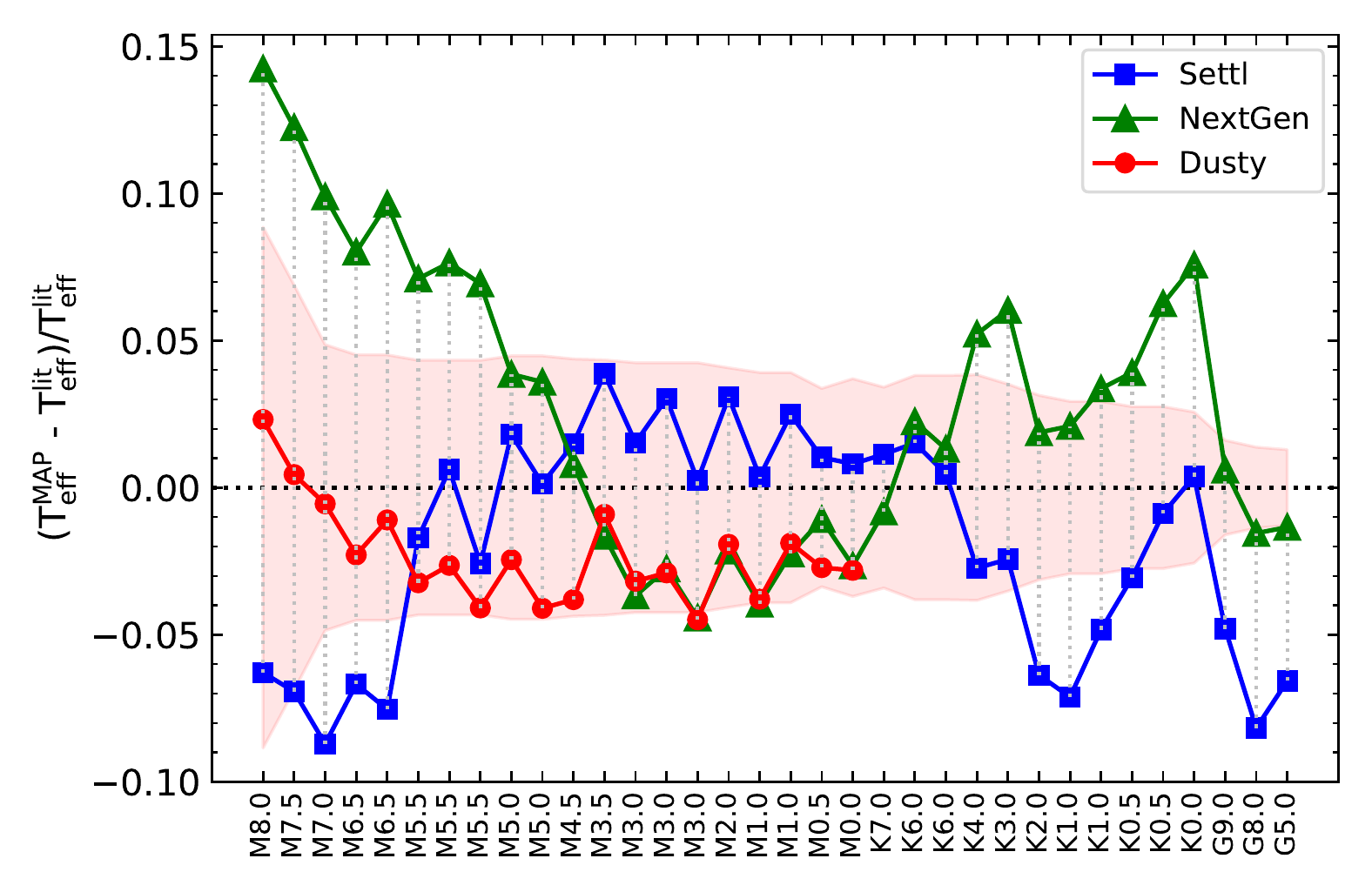}
    \caption{ Relative temperature deviations of the template stars between the MAP estimates and the literature values sorted by their spectral type. Different colours and symbols indicate the results of three different cINNs. The pink area indicates the uncertainty of the literature value of temperature. We only present template stars whose literature value of temperature is within the network training range.
     } \label{fig:tpl_dt_comp}
\end{figure}

To compare the performance of the three networks on the temperature in more detail, we present the relative temperature deviations between the MAP predictions and the literature values sorted by their spectral type. Figure~\ref{fig:tpl_dt_comp} as well shows that MAP estimates from \dust\ are in good agreement with the literature value within a 5 per cent discrepancy. In the case of \dust, all but one star have a deviation within the 1-subclass interval. In the case of \settl\ and \nextgen, 23 and 16 stars out of 34, respectively, have a deviation less than a 1-subclass interval. MAP estimates of \settl\ and \nextgen\ have a relatively poor agreement with the literature values for hot stars of $4500\,$K (e.g. K4.0 type) or higher. However, the discrepancies are still within 10 per cent.
The average absolute relative deviations when only using the templates within the training range of each network are 3.28, 4.49, and 2.58 per cent for \settl, \nextgen, and \dust, respectively (Table~\ref{table:avg_rerr_3net}). These average errors are equivalent to 1.08, 1.12, and 0.601 subclasses.

\begin{figure}
	\includegraphics[width=1\columnwidth]{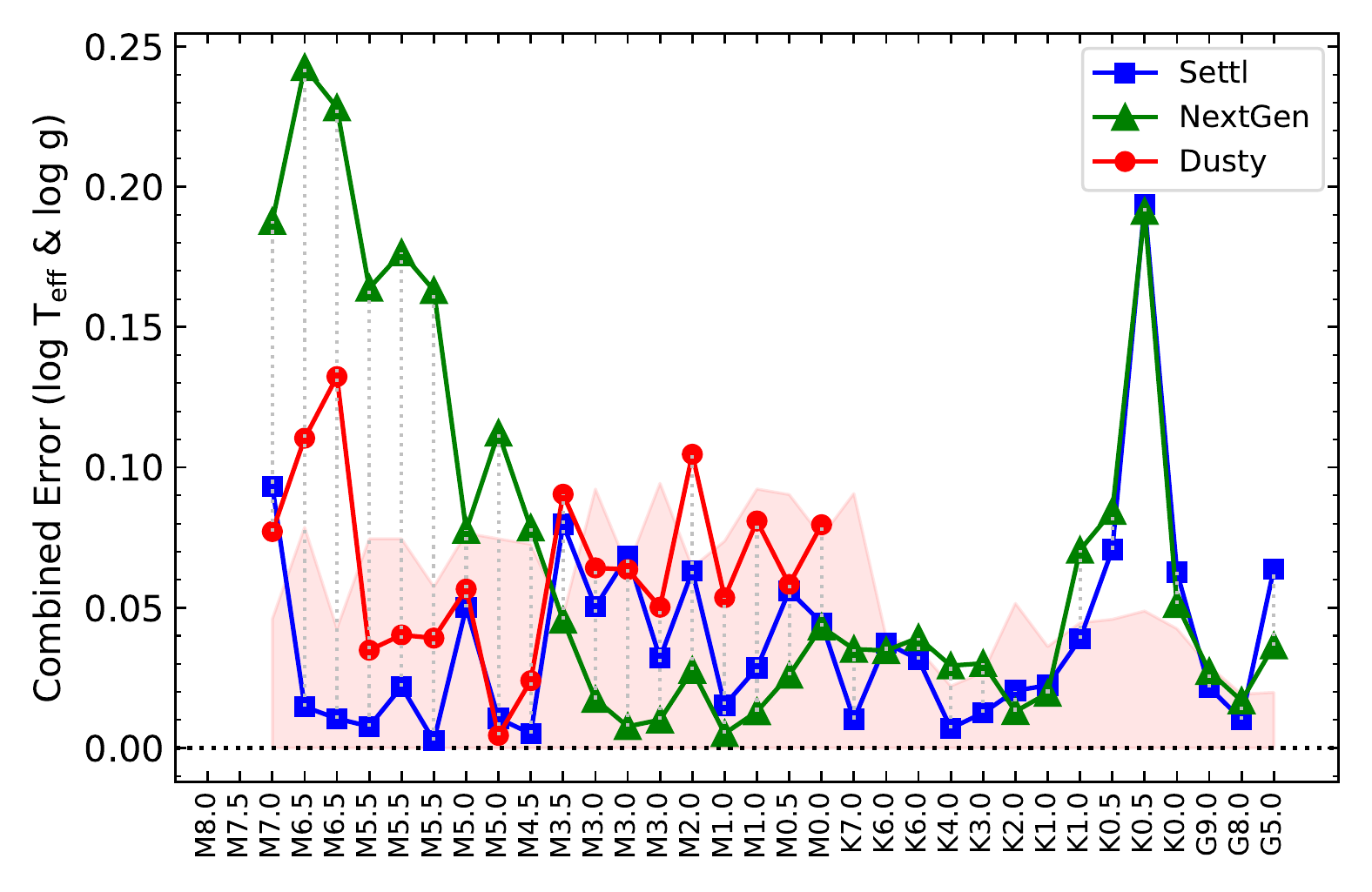}
    \caption{ Average relative error of the template stars between the MAP estimates and the literature values sorted by their spectral type. The average error is calculated as a root mean square of the relative errors of temperature and gravity, both in log-scale (Eq.~\ref{eq:cberr}). The pink area indicates the 1-$\sigma$ uncertainty of the literature value. We only present template stars whose literature value of temperature is within the network training range and whose literature value of gravity is presented. Colour codes are the same as in Fig.~\ref{fig:tpl_dt_comp}.
     } \label{fig:tpl_cerr_comp}
\end{figure}

\renewcommand{\arraystretch}{1.25}
\begin{table}
    \centering
    \caption{ Average absolute relative error between cINN predictions and literature values for template stars.
    \label{table:avg_rerr_3net}}
    \resizebox{0.48\textwidth}{!}{
    \begin{tabular}{ l  c  c  c   c c c}
        \toprule
        \multicolumn{1}{l}{} & \multicolumn{3}{c}{Average relative error [\%]} & \multicolumn{3}{c}{Average relative error [$\sigma$]} \\
        \cmidrule(rl){2-4} \cmidrule(rl){5-7}
        Network & $T_{\mathrm{eff}}$ & $\log(g)$ & $A_{\mathrm{V}}$ &
                $T_{\mathrm{eff}}$ & $\log(g)$ & $A_{\mathrm{V}}$ \\
                
        \midrule
            Settl & 3.28 & 5.5 & - & 1.08 & 0.809 & 2.78 \\  
            NextGen & 4.49 & 10.2 & - & 1.12 & 1.38 & 4.95 \\  
            Dusty & 2.58 & 9.13 & - & 0.601 & 1 & 3.87 \\  

        \bottomrule
    \end{tabular}} 
    \tablefoot{We calculate the errors by dividing the absolute difference between the MAP estimate and literature value either by literature values (i.e. errors in per cent unit) or by 1-$\sigma$ uncertainty of literature value (i.e. errors in 1-$\sigma$ unit). In the case of the effective temperature, the 1-$\sigma$ uncertainty corresponds to the temperature interval of one subclass. For each network and parameter, we only use template stars whose literature values are within the training range of the network to calculate the errors.}
\end{table}

\begin{figure*}
    \centering
    \includegraphics[width=2\columnwidth]{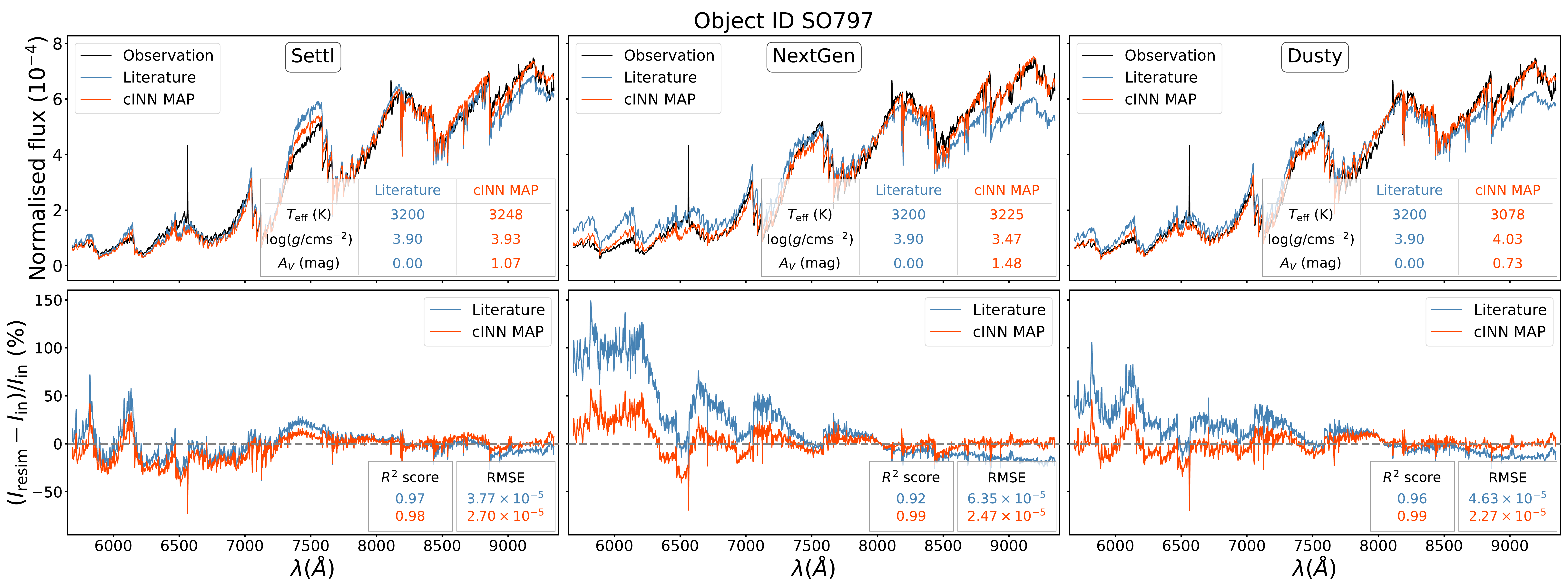}
    \caption{Resimulation results for Class III star SO797. The columns show in order the results for the three different spectral libraries Settl, NextGen and Dusty. Top: Comparison of resimulated spectrum. The blue spectrum indicates the resimulation derived from the literature stellar parameters from Table~\ref{tab:class_III_templates}. The red spectrum shows the corresponding resimulation based on the cINN MAP prediction. The respective input parameters for the resimulation are summarised in the table in the bottom right corner. The relative residuals $(I_\mathrm{resim} - I_\mathrm{in}) / I_\mathrm{in}$ of the resimulated spectra compared to the input spectrum are shown in the bottom panels, respectively.}
    \label{fig:ResimTemplates_Spectrum_ID_SO797}
\end{figure*}

In the case of surface gravity, the RMSE of \settl, \nextgen, and \dust\ are 0.30, 0.51, and 0.42~dex, respectively. However, because the surface gravity value from previous studies~\citep{Stelzer2013, Manara2017} is obtained by fitting the spectrum on the Settl models, the MAP estimate of \settl\ is essentially the closest to the literature value. Although \settl\ has the smallest RMSE value, considering the uncertainty of literature values, the other two networks also have good agreement with the literature value.

To combine the results of temperature and surface gravity, we define the combined error of two parameters as,
\begin{equation}
\label{eq:cberr}
\begin{split}
    \mathrm{Combined \: error} &=  \sqrt{ \frac{1}{2} \left( \left( \frac{ \Delta T_{\mathrm{eff}} }{\mathrm{log} \,  T^{\mathrm{lit}}_{\mathrm{eff}}}\right)^{2}  
   +  \left( \frac{ \Delta g }{ \mathrm{log} \, g^{\mathrm{lit}}}\right)^{2} \right) }  ,  \\
   \: \mathrm{for} \\
   \Delta T_{\mathrm{eff}} &= \mathrm{log} \, T^{\mathrm{MAP}}_{\mathrm{eff}} - \mathrm{log} \, T^{\mathrm{lit}}_{\mathrm{eff}}, \\
   \Delta g &=  \mathrm{log} \, g^{\mathrm{MAP}} - \mathrm{log} \, g^{\mathrm{lit}},
\end{split}
\end{equation}
and present the combined error of each template star. We use the effective temperature in the logarithmic scale to match the scale with surface gravity. The overall result using combined error presented in Fig.~\ref{fig:tpl_cerr_comp} are not significantly different from Fig.~\ref{fig:tpl_dt_comp}, but by adding the gravity error, \settl\ shows better performance than \dust\ even for low-temperature stars. In the case of \nextgen, the combined error is larger than the other two networks because there are cases where temperature and gravity errors are both large.
The average combined errors across the stars of \settl\, \nextgen, and \dust\ are 3.93, 7.20, and 6.47 per cent, respectively.

In the case of \settl, all but 7 stars are in good agreement with the literature values within 1-$\sigma$ uncertainty. Excluding one star with a large error, most of the stars have errors of less than 5 per cent and a maximum of 10 per cent. \dust\ also has small errors ($<$15 per cent) but \dust\ has a disadvantage in that it is inherently less versatile than the other two networks because of its training range. \nextgen\ also shows an error of less than 10 per cent for stars with spectral type earlier than M5.0.

Lastly, in the case of extinction, the deviation between MAP estimates and literature values varies depending on the temperature. For stars hotter than about 3400~K (i.e. M3.0 type), all three networks predict near-zero extinction, with little deviation from literature values. In the case of \nextgen, there are stars that are slightly outside the error range but their MAP estimates are sufficiently small. On the other hand, for cool stars below 3400~K, the discrepancy between the MAP value and the literature value gradually grows. In the case of \settl\ and \dust, the MAP estimate does not exceed the maximum of 3, but in the case of \nextgen, the MAP estimates are slightly larger than the other two networks.

In this section, we showed that the discrepancy between the network MAP prediction and literature value varies with the characteristics of the stars. Based on the overall results, a star of 
\begin{itemize}
    \item M6.5 -- K1.0 (2935 -- 5000~K) for \settl,
    \item M4.5 -- K1.0 (3200 -- 5000~K) for \nextgen, 
    \item M5.5 -- M0.0 (3060 -- 4000~K) for \dust\ 
\end{itemize}
shows especially high agreement with the literature values.
\settl\ showed the best agreement with the literature values overall. \dust\ also shows a good agreement for stars whose temperature is within the Dusty database of 2600 -- 4000~K. \nextgen\ has relatively large errors compared to the other two, but it still shows reliable performance for early-type stars.
Given that \settl\ and \nextgen\ cover a wider range of temperature (i.e. 2600 -- 7000~K) and gravity (2.5 -- 5 $\log(\mathrm{cm\,s}^{-2})$) than \dust, \settl\ is the best choice among the three networks. However, all three networks showed good agreement with the literature values considering their uncertainty.

This result shows how well our cINN predictions are in good agreement with values obtained with the classical methods in previous studies. Differences between literature values and network predictions do not demonstrate that the network prediction is wrong. For example, in the case of surface gravity, because the literature value was also obtained by fitting spectra based on the Settl model, there is inevitably a larger discrepancy between the literature values and MAP predictions of \nextgen\ and \dust. This means that we need to consider methods used in the literature, and additional analysis is required to judge whether the cINN prediction is really wrong or not. The resimulation following in the next section will provide a better clue to determine the correctness of our cINN predictions.

\subsubsection{Resimulation}
\label{sec:ResimTemplates}

\begin{figure}
    \centering
    \includegraphics[width=1\columnwidth]{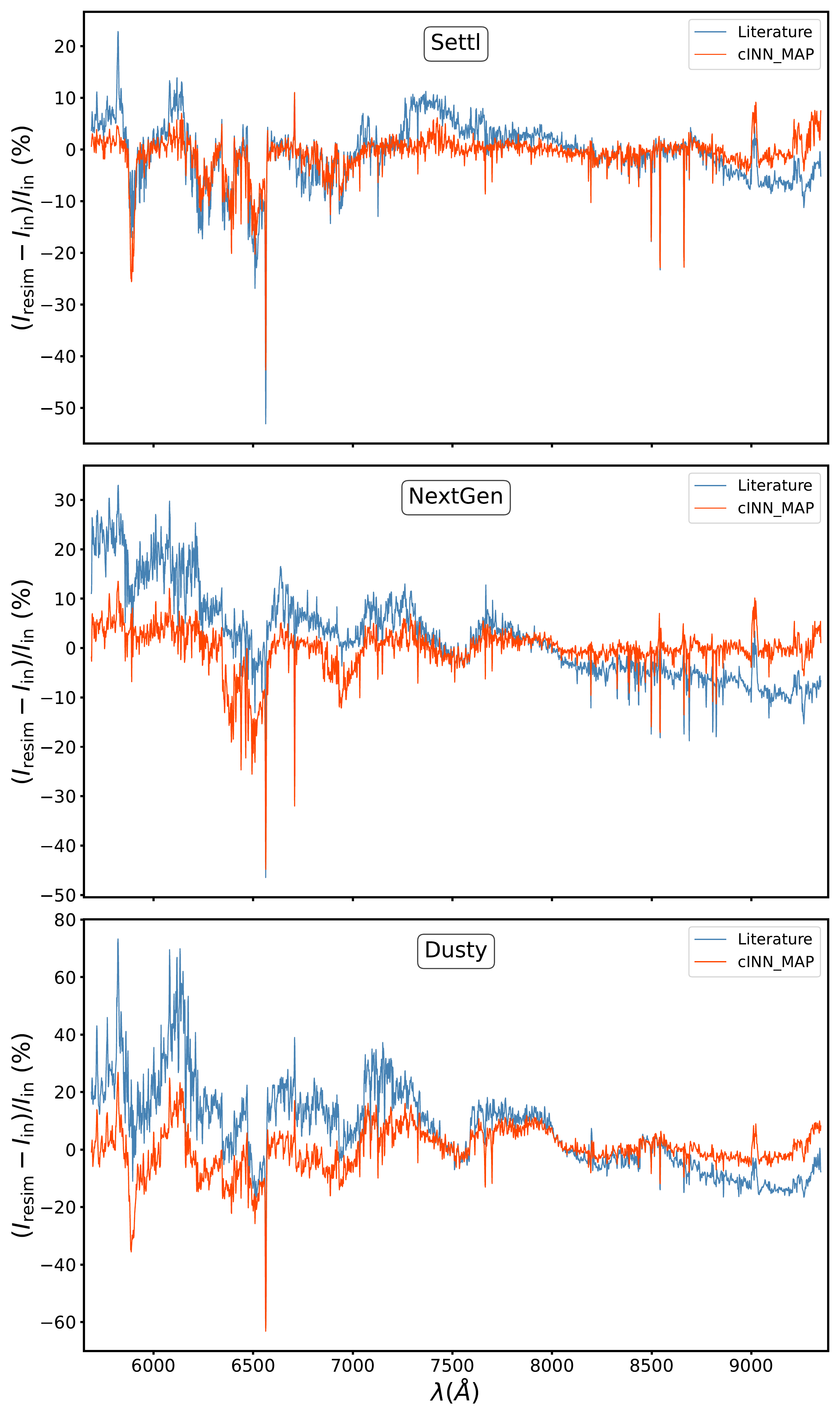}
    \caption{Comparison of the median relative error of the resimulated spectra for the Class III template stars between the resimulations based on the literature stellar parameters (blue, see Table~\ref{tab:class_III_templates}) and the cINN MAP predictions (red). From top to bottom, the panels show the corresponding results for the three tested spectral libraries Settl, NextGen and Dusty.}
    \label{fig:ResimTemplates_MedErr_ModelComparison}
\end{figure}

As we have done for the synthetic test data in Sect. \ref{sec:ResimSynth}, we also evaluate the accuracy of the cINN predictions on the Class III template by resimulation to quantify the agreement between the spectra corresponding to the MAP estimates with the input spectra. In this case, we also run a resimulation for the nominal literature stellar parameters of the Class III sources listed in Table~\ref{tab:class_III_templates} for comparison. Some of the Class III template sources in our sample do not have an estimate for $\log(g)$ in the literature. For these sources, we assume a fixed value of $\log(g/\mathrm{cms}^{-2}) = 4.0$ in our resimulation, which is a reasonable guess for the spectral types in our sample. The sources in question are marked as "fixed" in the last column of Table~\ref{tab:class_III_templates}. There are a few templates (7 for Settl, 1 for NextGen and 8 for Dusty; see Table~\ref{tab:SummaryTemplateMAP}), where the cINN extinction MAP estimate has a non-physical negative value. Since most of these are only barely below zero, we decide to allow these negative values to be accounted for during the resimulation. 

Figure~\ref{fig:ResimTemplates_Spectrum_ID_SO797} shows an example result of the resimulation for the M4-type template star SO797 for all three spectral libraries with the top panels comparing the resimulated spectra to the input spectrum and bottom panels showing the corresponding residuals. Here the red curve indicates the resimulation result derived from the cINN MAP estimates, whereas the blue curve marks the literature-based outcome. In this particular example, the cINN recovers both $T_\mathrm{eff}$ and $\log(g)$ quite accurately for all three spectral libraries but overestimates $A_V$ for this supposedly zero extinction template Class III source by $1.07$, $1.48$ and $0.73$ mag based on Settl, NextGen and Dusty, respectively. Interestingly, however, we find that the resimulated spectrum based on the cINN MAP prediction with the supposedly wrong $A_{\mathrm{V}}$ matches the input spectrum better than the spectrum derived from the literature value in all three examples as e.g.~attested by the smaller RMSE and better $R^2$ score of $2.7 \times 10^{-5}$ and $0.98$ compared to $3.77 \times 10^{-5}$ and $0.97$ in the Settl case. Figure~\ref{fig:ResimTemplates_Spectrum_ID_RXJ0445.8+1556} in the Appendix shows another such example, where it is immediately apparent that the cINN-based resimulated spectrum matches the input observation much better than the literature-based solution, which evidently does not capture the slope of the observed spectrum correctly.

Figures~\ref{fig:ResimTemplates_MedErr_ModelComparison}, \ref{fig:ResimTemplates_RMSE_histogramm} and Table~\ref{tab:ResimTemplatesRMSEs} in the Appendix summarise the resimulation results across the entire Class III template sample, showing the median relative residuals against the wavelength, the distributions of RMSEs and $R^2$ scores, and a table of all RMSEs and $R^2$ scores, respectively. Note that the resimulation statistics vary between the libraries here. Given the lower effective temperature limits of the libraries (i.e.~$2600$ K) 2 of the 36 templates, namely TWA26 and DENIS1245, can a priori not be resimulated with Settl and NextGen. For Dusty, the literature sample is even smaller with only 20 out of 36 templates due to the low upper temperature limit of 4000 K. For the resimulation of the MAP estimates we can use 31 templates with the \settl, 29 with \nextgen\ and only 17 with \dust. For more details, we refer to Table~\ref{tab:ResimTemplatesRMSEs}. Note that for the Dusty resimulation there are actually 7 templates, where the $\log(g)$ prediction is above the training set limit of 5. However, since the Dusty spectral library does actually extend to $\log(g/\mathrm{cms}^{-2}) = 5.5$, we decide to run the resimulation for these 7 templates anyways, in particular since for most of those the $\log(g)$ prediction is only barely above 5 (see Table~\ref{tab:SummaryTemplateMAP}). 

Figure~\ref{fig:ResimTemplates_MedErr_ModelComparison} shows that for all three libraries, we find that our observation from Fig.~\ref{fig:ResimTemplates_Spectrum_ID_SO797}, where the resimulated spectrum based on the cINN prediction fits the input spectrum better than the literature-based resimulation, holds on average across the entire template sample. The distributions of the RMSEs and $R^2$ scores of the resimulated spectra in Fig.~\ref{fig:ResimTemplates_RMSE_histogramm} further confirm this, as the cINN-based resimulated spectra tend towards smaller RMSEs and slightly better $R^2$ scores compared to the literature-based spectra for all three spectral libraries.

Examining the 7 templates, for which the Dusty-based cINN prediction of $\log(g)$ exceeds the learned upper limit of 5 (i.e.~the cINN extrapolated), more closely, the resimulation results show that even when the cINN extrapolates, the set of predicted parameters corresponds to a spectrum, which matches the input observation quite well and, in particular, equally if not better than the respective spectrum resimulated from the literature values as indicated by the $R^2$ scores (see Table~\ref{tab:ResimTemplatesRMSEs} and Fig.~\ref{fig:ResimTemplates_Spectrum_ID_CD_29_8887A} for an example). This result shows that the cINN prediction is actually fairly robust even in the event of slight extrapolation.

\begin{figure*}
    \centering
    \includegraphics[width=0.99\textwidth]{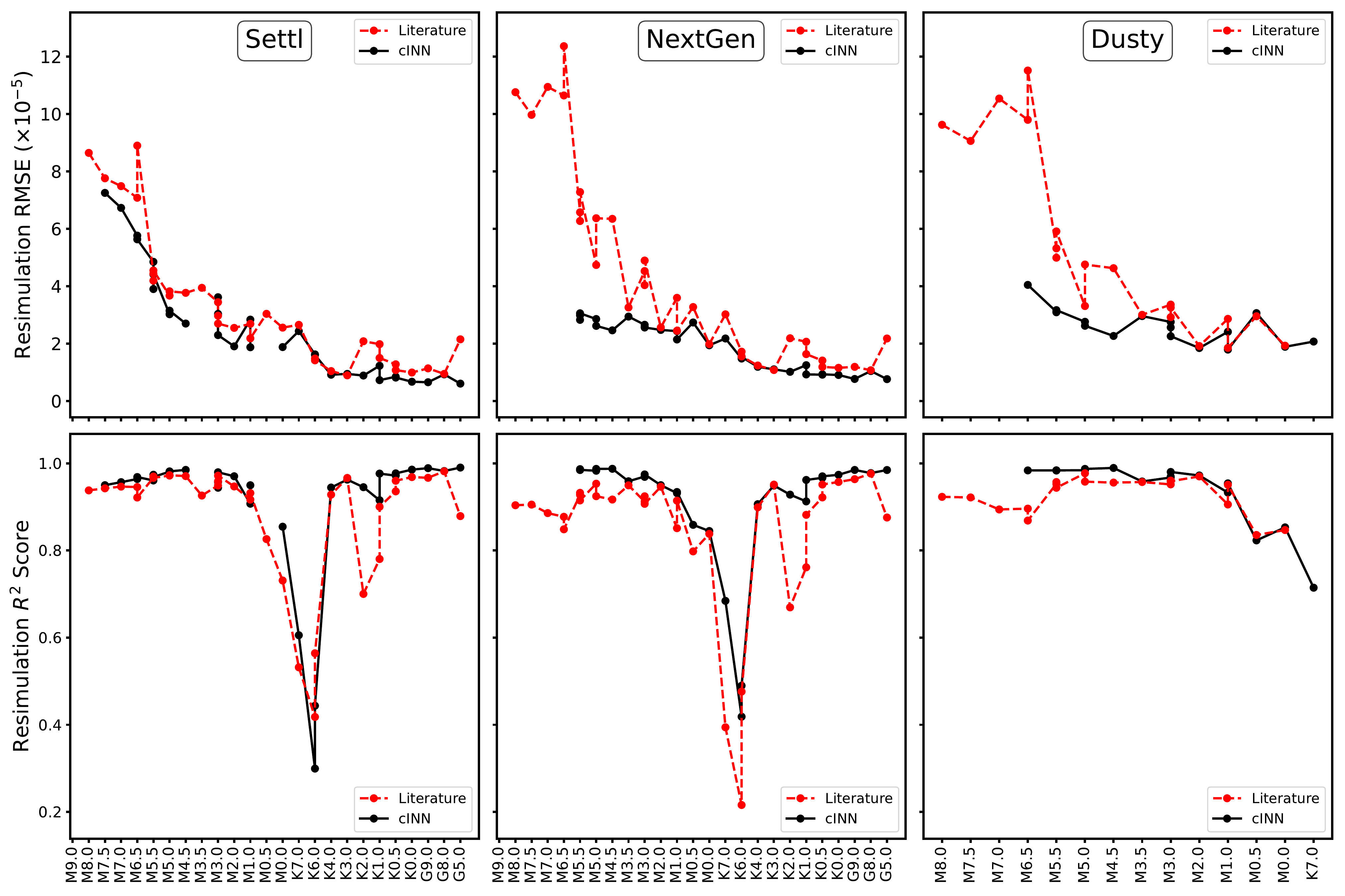}
    \caption{Comparison of the resimulation accuracy measures (RMSE in the top row, $R^2$ score in the bottom) for the three spectra libraries to the spectral type of the Class III templates. In all panels, the dotted red line indicates the results for the resimulation based on the literature stellar properties, while the black line shows the cINN-based outcomes.}
    \label{fig:Resim_RMSE_R2_vs_SpT}
\end{figure*}

Comparing our chosen resimulation accuracy measures to the spectral types of the Class III templates in Fig.~\ref{fig:Resim_RMSE_R2_vs_SpT}, we find that the RMSEs exhibit an increasing trend towards the M-types for all three spectral libraries. For the $R^2$ scores, we find a notable dip in the goodness of fit for the intermediate spectral types, that is between M2 to K3, in both the resimulation of the literature and cINN-based values for Settl and NextGen. The beginning of this dip can also be seen in the Dusty-based results up to the temperature limit of this library at the K7 type. Interestingly, when compared to Fig.~\ref{fig:tpl_cerr_comp} in this spectral type range the discrepancy between the cINN prediction and literature stellar properties is relatively low, where both cINN and literature values correspond to an equally sub-optimal fit to the observed spectra. 

Overall the resimulation test shows that the cINN approach predicts parameters for the real Class III template spectra that correspond to spectra, which not only fit the input observations very well (as shown by the good $R^2$ scores in Fig.~\ref{fig:ResimTemplates_RMSE_histogramm} and Table~\ref{tab:ResimTemplatesRMSEs}), but also match better than the spectra resimulated from the literature values in most instances.

\begin{figure*}
    \centering
    \includegraphics[width=0.99\textwidth]{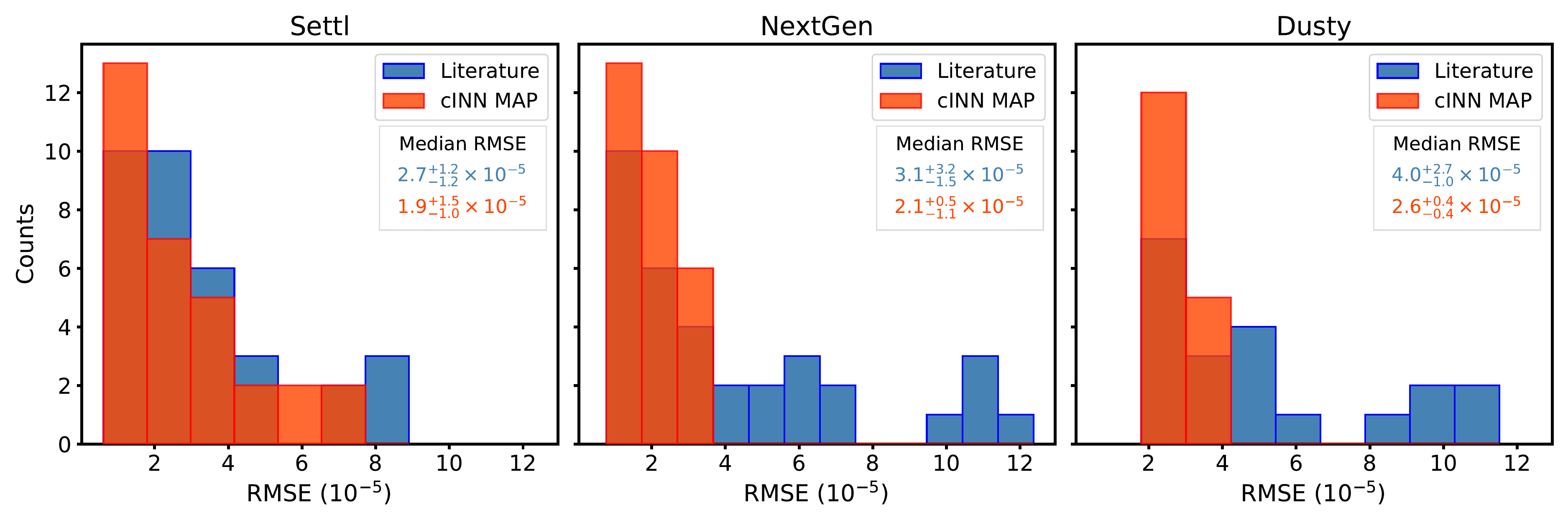}
    \includegraphics[width=0.99\textwidth]{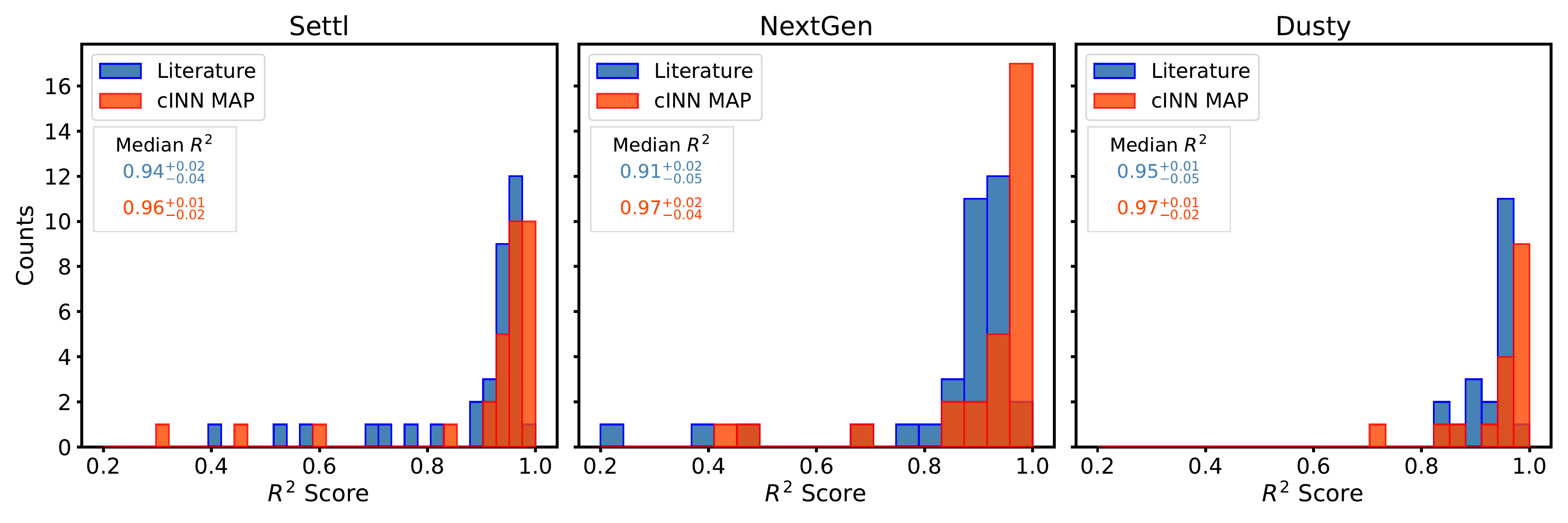}
    \caption{Average error for the resimulation spectra for the Class III template stars.
    Top: Histograms of the RMSEs for the resimulation on the Class III template spectra for the three different spectral libraries. Bottom: Histograms of the corresponding $R^2$ scores for the resimulated spectra.}
    \label{fig:ResimTemplates_RMSE_histogramm}
\end{figure*}

\section{Feature importance}
\label{sec:featureimportance}

\subsection{Importance calculation}
In this section, we evaluate which parts of the spectra the cINN prediction relies the most upon. To do so we measure the so-called \textit{permutation feature importance}, an approach first described by \cite{Breiman2001} for random forest models and later generalised by \cite{FischerRudinDominici2018}. In this study we implement the \cite{FischerRudinDominici2018} algorithm as described in \cite{Molnar2022}, operating as follows:

First, we compute the error on the original held-out test set
\begin{equation}
    e_\mathrm{orig} = L\left(\mathbf{X}, g(\mathbf{Y})\right),
\end{equation}
where $g$ represents the inverse translation ($\mathbf{x} \leftarrow \mathbf{y}$) of the trained cINN, $\mathbf{X}$ denotes the matrix of the target parameters of the test set ($n_\mathrm{test} \times n_\mathrm{parameters}$), $\mathbf{Y}$ is the $n_\mathrm{test} \times n_\mathrm{features}$ feature matrix of the test set and $L$ represents a loss measure. In our case, $L$ is the RMSE of the MAP estimates.

Next, for each feature $j \in \{1, \ldots, n_\mathrm{features}\}$, we generate a feature matrix $\mathbf{Y}_{\mathrm{perm}, j}$ via random permutation of the $j$-th column in order to break the association between feature $j$ and the target parameters $\mathbf{x}$, estimate the prediction error $e_{\mathrm{perm}, j} = L\left(\mathbf{X}, g\left(\mathbf{Y}_{\mathrm{perm}, j}\right)\right)$ based on the permuted data set, and compute the feature importance of feature $j$ as the quotient 
    \begin{equation}
        \mathrm{FI}_j = \frac{e_{\mathrm{perm}, j}}{e_\mathrm{orig}}.
    \end{equation}
The larger $\mathrm{FI}_j$ is, the worse the model prediction becomes if feature $j$ is scrambled via permutation, that is the more important feature $j$ is to the model's decision making. The closer $\mathrm{FI}_j$ is to 1, on the other hand, the less feature $j$ affects the predictive performance and, thus, the less relevant it is to the model's reasoning.

In our particular case, the feature space is very high dimensional with $2930$ spectral bins per spectrum. Consequently, computing the individual per spectral bin feature importance is rather computationally expensive as it requires generating the posteriors and determining the MAP estimates for each of the $2930$ bins. Although the computational cost alone is not prohibitive in this case given the cINNs great efficiency, we still opt for a slightly different approach, because the spectral bins themselves are also not necessarily independent of each other. Instead of using the individual bins, we group them together into combined features, for which we then estimate the importance. In practise, this means that we permute multiple columns at once (each column with its own permutation seed though) corresponding to the spectral bins in a given group. For the setup in this study in particular we decide to evaluate the feature importance across the wavelength range using groups of 10 bins, which corresponds to a spectral width of $12.5$ \AA. We set all groups to overlap by 5 bins (i.e. $6.25$ \AA) with the preceding and following groups. We average feature importance for overlapping bins.

\subsection{Important features for M-, K-, and G-type stars}

\begin{figure*}
	\includegraphics[width=2\columnwidth]{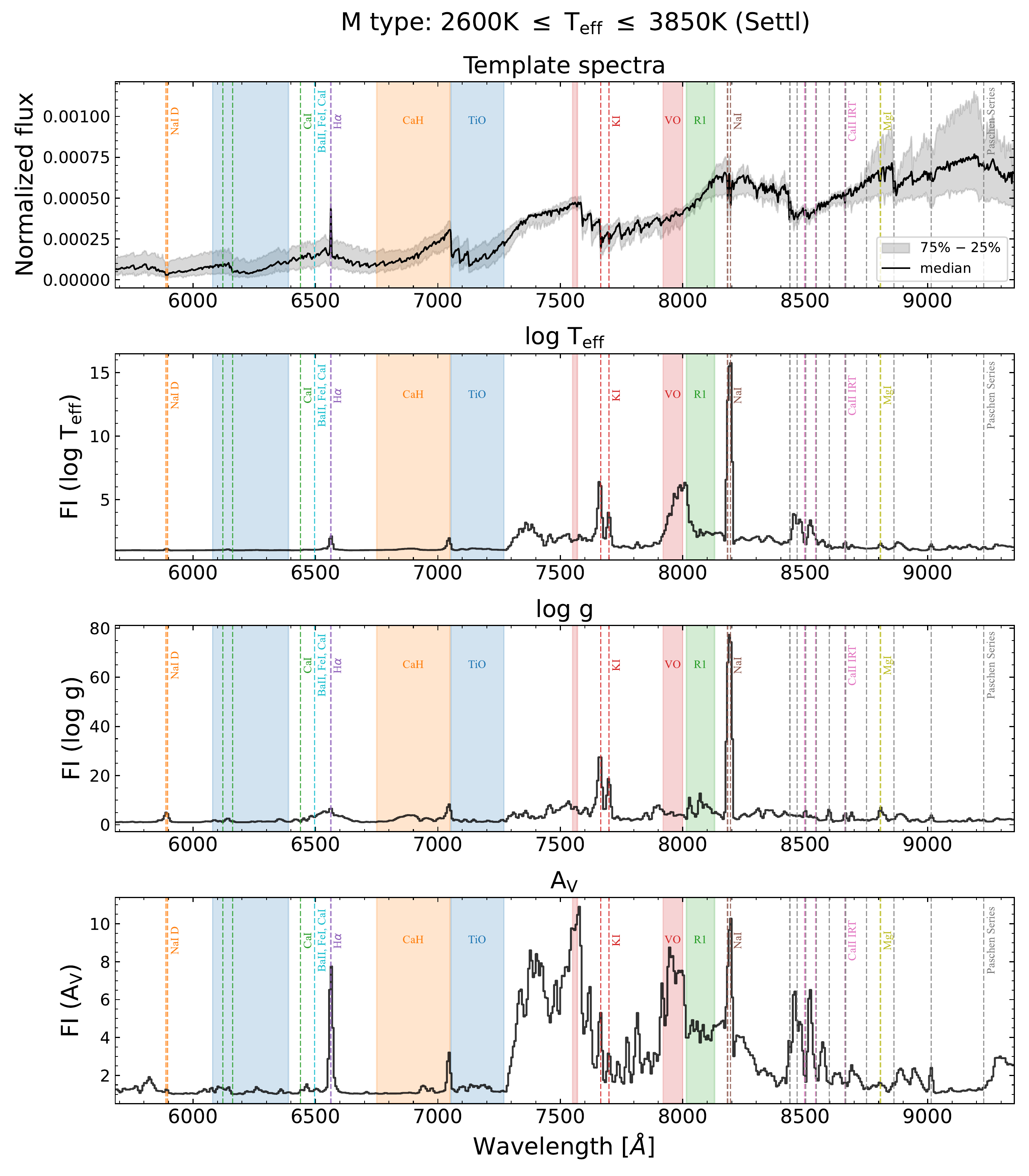}
    \caption{Feature importance evaluation for M-type synthetic models in the test set using \settl. We present the median flux of M-type Class III template stars in the first row. The grey area indicates the interquantile range between the 25\%\ and 75\%\ quantiles. The other three rows show the feature importance across the wavelength for each stellar parameter. Vertical lines and shades indicate the location of typical tracers of stellar parameters listed in Table~\ref{table:fi_tracer}.
     } \label{fig:fi_stl_m}
\end{figure*}

\begin{figure*}
	\includegraphics[width=1\columnwidth]{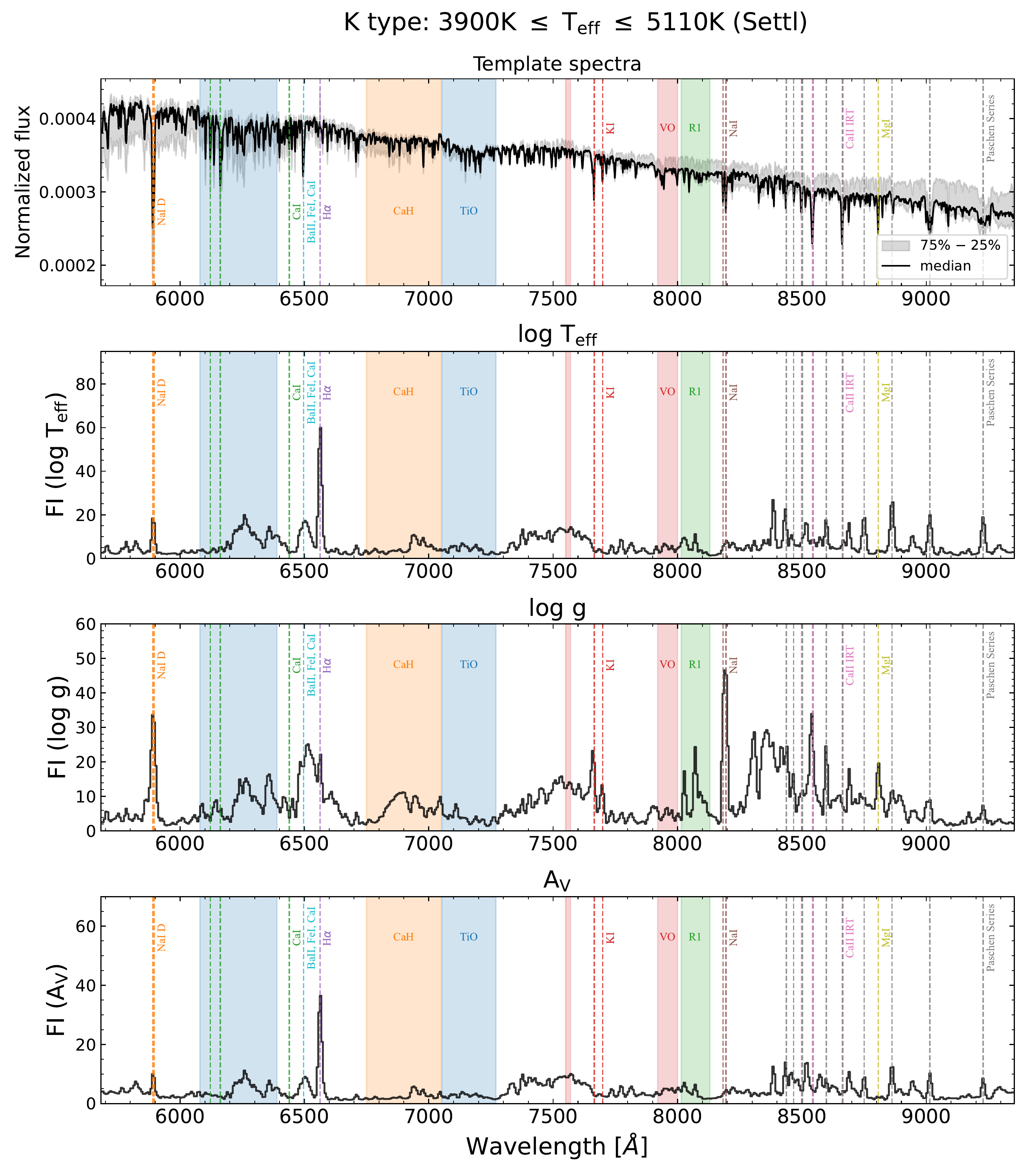}
	\includegraphics[width=1\columnwidth]{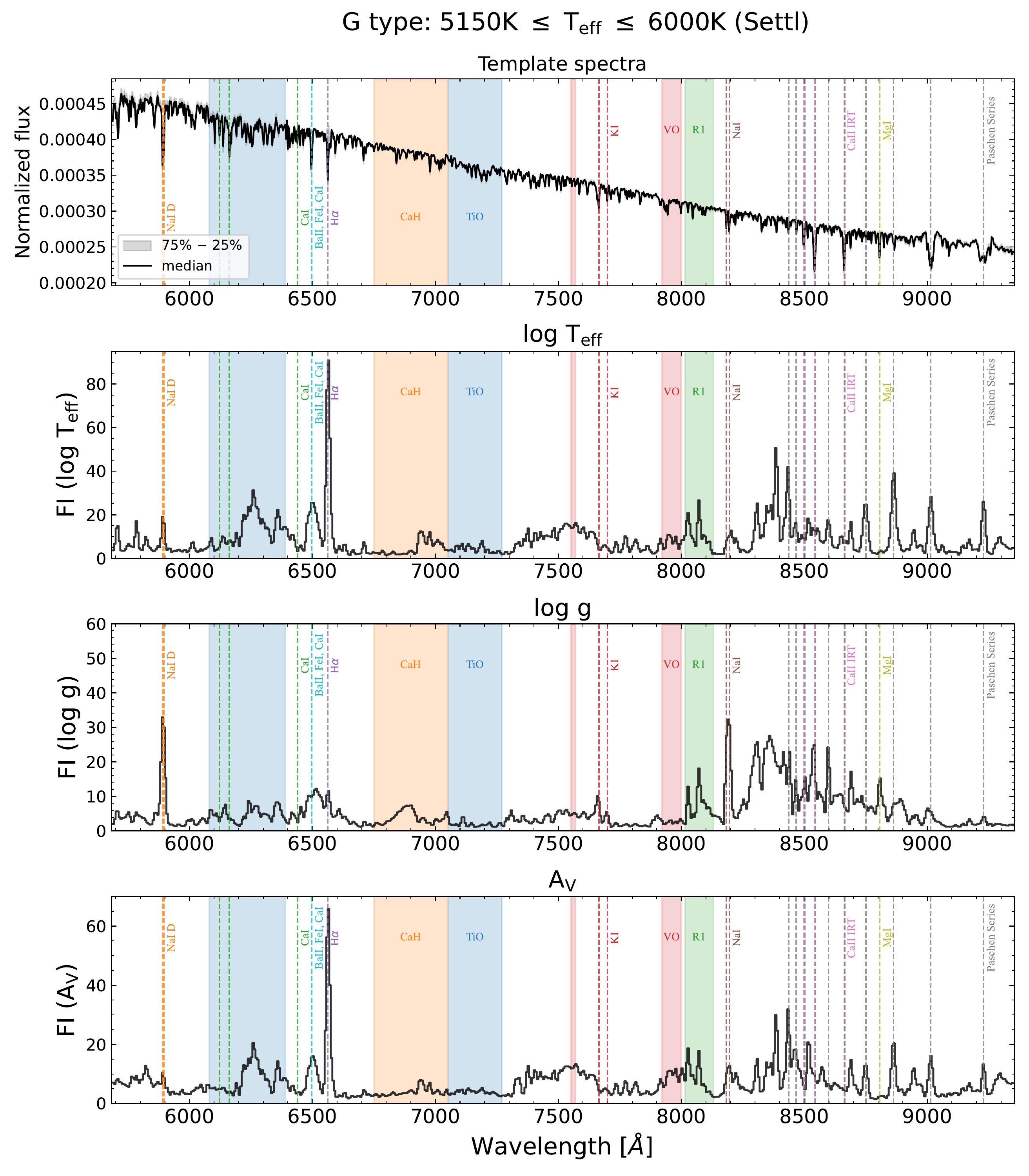}
    \caption{Feature importance evaluation for K-type synthetic models (left) and for G-type synthetic models (right) in the test set using \settl.
    The panels in the first row show the median flux of K-type and G-type Class III template stars, respectively. 
    Lines and shades are the same as Fig.~\ref{fig:fi_stl_m}.
     } \label{fig:fi_stl_kg}
\end{figure*}

We draw three groups from the test set according to the temperature of the test model: M-type (2600K--3850~K) group, K-type (3900K--5110~K) group, and G-type (5150K--6000~K) group, and evaluate the feature importance across the wavelength for each group per network. In the case of \dust, we only evaluate for the M-type group because the maximum temperature of the Dusty database is 4000~K.

Figure~\ref{fig:fi_stl_m} presents the feature importance of \settl\ for M-type stars. To compare the important features with the locations of stellar parameter tracers existing in the real spectrum, we plot the median flux of M-type template stars in the first row, and indicate the locations of several tracers of stellar parameters (Table~\ref{table:fi_tracer}): \nai\ doublet $5890$, $5896$~\AA~\citep[\teff\ and \logg,][]{Allen1995}, \cai\ $6122$, $6162$, $6439$~\AA~\citep[\logg,][]{Allen1995}, \baii, \fei, and \cai\ blend 6497~\AA~\citep[\teff\ and \logg,][]{Allen1995, Herczeg2014}, \ha\ $6563$~\AA~\citep[\teff,][]{Luhman2003}, \ki\ doublet $7665$, $7699$~\AA~\citep[\teff\ and \logg,][]{Manara2013,Manara2017}, \nai\ doublet $8183$, $8195$~\AA~\citep[\teff\ and \logg,][]{Kirkpatrick1991, Allen1995, Riddick2007}, \caii\ IR triplet $8498$, $8542$, $8662$~\AA~\citep[\teff,][]{Kirkpatrick1991, Allen1995, Luhman2003}, \mgi\ $8807$~\AA~\citep[\teff,][]{Manara2013,Herczeg2014}, hydrogen Paschen series~\citep[\av,][]{Edwards2013}, CaH $6750$--$7050$~\AA~\citep[\teff\ and \logg,][]{Kirkpatrick1993, Allen1995}, TiO $6080$--$6390$, $7053$--$7270$~\AA~\citep[\teff,][]{Kirkpatrick1991,Henry1994, Jeffries2007}, ViO $7550$--$7570$, $7920$--$8000$~\AA~\citep[\teff,][]{Allen1995, Riddick2007, Manara2013}, and R1 $8015$--$8130$~\AA~\citep[\teff,][]{Riddick2007} .

To evaluate whether these observational tracers act as important features in our networks, we check whether the feature importance value corresponding to each tracer's wavelength is larger than a fiducial value. We use the value of median plus one standard deviation over the entire wavelength range as a fiducial value to determine an important tracer. For tracers with multiple lines or molecular bands, we average the feature importance for each line or over the wavelength range. In Table~\ref{table:fi_tracer}, we mark tracers whose average importance is larger than the fiducial value. We also indicate for which parameters these lines and bands trace in real observations.

Figure~\ref{fig:fi_stl_m} shows that \nai\ doublet $8183$, $8195$~\AA\ lines are the most important feature for \settl\ to predict stellar parameters of M-type stars. In the case of extinction, there are two wide peaks near 7500~\AA, where the redder peak overlaps with the VO molecular band. However, \nai\ has a similarly large importance value. In the case of temperature and gravity, \ki\ doublet $7665$, $7699$~\AA\ lines play a second important role, and in extinction, \ha\ does. VO and R1 molecular absorption bands as well act as important features to determine the temperature and extinction.

We present the feature importance evaluated for \nextgen\ and \dust\ in Fig.~\ref{fig:fi_ng_m}. The fact that \nai, \ki, and \ha\ are important features for M-type stars is the same for all three networks. However, for \nextgen, there is a large bump at 7500~\AA\ in the case of temperature. The results of \nextgen\ in overall are spikier than in the other two networks. 
In the case of \dust, the importance value of \nai\ doublet $5890$, $5896$~\AA\ (\nai\ D) is relatively large compared to the other networks, and there is a very wide bump around \nai\ doublet $8183$, $8195$\AA.

Given the fact that extinction affects the overall shape of the spectrum, it is interesting that the \settl\ relies a lot on a few certain lines. Broad bumps exist in the red part of the spectrum, but there are particularly important lines and areas such as the \nai, \ha, and near VO bands. The result of \nextgen\ is similar to \settl\ but shows a little more spiky trend with wider peaks. \dust\ shows a more wavy shape across the entire wavelength range compared to the others.

Next, in the case of K-type stars, the results of \settl\ and \nextgen\ are similar to each other, unlike the case of M-type stars, so we only present the result of \settl\ in this paper (left panels in Fig.~\ref{fig:fi_stl_kg}).
Compared to the results of M-type stars, it is noticeable that important features are different for each parameter. In the case of temperature and extinction, the overall shapes are similar, with the \ha\ line being the most important feature. \nai\ doublet $8183$, $8195$~\AA\ are no longer so important to determine temperature and extinction for K-type stars. In addition, \nai\ D lines and hydrogen Paschen series have relatively high importance values. 
On the other hand, in the case of surface gravity, the \nai\ doublet $8183$, $8195$~\AA\ lines still play the most important role. The importance of \nai\ D in gravity becomes noticeable in K-type stars compared to M-type stars. Additionally, there are several peaks at \ki, \mgi\ $8807$~\AA\ used as important features to determine gravity.

The result of G-type stars (i.e. right panels in Fig.~\ref{fig:fi_stl_kg}) is similar to the K-type stars. The \ha\ is still the most important feature for temperature and extinction and there are several peaks at the Paschen series as well. For gravity, \nai\ D becomes more important in G-type stars and has an importance value comparable to that of \nai\ doublet $8183$, $8195$\AA. These sodium lines are the most important features to determine gravity. On the other hand, the importance of \ki\ lines decreases in G-type stars compared to K-type stars.

\renewcommand{\arraystretch}{1.25}
\begin{table*}
    \caption{We mark tracers whose feature importance values are larger than the fiducial value of median plus 1 standard deviation, meaning that marked tracers are significantly important features to determine each stellar parameter.
    \label{table:fi_tracer}}
    \begin{tabular}{ l l  c  c  c   c c c  c c c}
        \toprule
        \multicolumn{1}{l}{} & \multicolumn{1}{l}{} & \multicolumn{3}{c}{ M-type } & \multicolumn{3}{c}{ K-type } & \multicolumn{3}{c}{G-type} \\
        \cmidrule(rl){3-5} \cmidrule(rl){6-8} \cmidrule(rl){9-11}
        Tracers & used in observations for & $T_{\mathrm{eff}}$ & $\log(g)$ & $A_{\mathrm{V}}$ &
                $T_{\mathrm{eff}}$ & $\log(g)$ & $A_{\mathrm{V}}$ & 
                $T_{\mathrm{eff}}$ & $\log(g)$ & $A_{\mathrm{V}}$ \\
        \midrule
        
        \nai\ doublet $5890$, $5896$~\AA\ & $T_{\mathrm{eff}}$, $\log(g)$ & - & - & - & \checkmark  & \checkmark  & \checkmark  & \checkmark  & \checkmark  & \checkmark  \\ 
TiO $6080$--$6390$, $7053$--$7270$~\AA\ & $T_{\mathrm{eff}}$ (M- and late K-type) & - & - & - & - & - & - & - & - & - \\ 
\cai\ $6122$, $6162$, $6439$~\AA\ & $\log(g)$ & - & - & - & - & - & - & - & - & - \\ 
\baii, \fei, and \cai\ blend 6497~\AA\ & $T_{\mathrm{eff}}$, $\log(g)$ & - & - & - & \checkmark  & - & \checkmark  & \checkmark  & - & \checkmark  \\ 
\ha\ $6563$~\AA\ & $T_{\mathrm{eff}}$ (early type) & - & - & \checkmark  & \checkmark  & \checkmark  & \checkmark  & \checkmark  & \checkmark  & \checkmark  \\ 
CaH $6750$--$7050$~\AA\ & $T_{\mathrm{eff}}$ (M-type), $\log(g)$ & - & - & - & - & - & - & - & - & - \\ 
VO $7550$--$7570$, $7920$--$8000$~\AA\ & $T_{\mathrm{eff}}$ (M-type) & \checkmark  & \checkmark  & \checkmark  & \checkmark  & \checkmark  & \checkmark  & \checkmark  & - & \checkmark  \\ 
\ki\ doublet $7665$, $7699$~\AA\ & $T_{\mathrm{eff}}$, $\log(g)$ & \checkmark  & \checkmark  & \checkmark  & - & - & - & - & - & - \\ 
R1 $8015$--$8130$~\AA\ & $T_{\mathrm{eff}}$ (M-type) & \checkmark  & \checkmark  & \checkmark  & - & - & - & \checkmark  & \checkmark  & - \\ 
\nai\ doublet $8183$, $8195$~\AA\ & $T_{\mathrm{eff}}$ (M-type), $\log(g)$ & \checkmark  & \checkmark  & \checkmark  & \checkmark  & \checkmark  & - & - & \checkmark  & \checkmark  \\ 
hydrogen Paschen series & $A_{\mathrm{V}}$ & - & - & - & \checkmark  & \checkmark  & \checkmark  & \checkmark  & \checkmark  & \checkmark  \\ 
\caii\ IR triplet $8498$, $8542$, $8662$~\AA\ & $T_{\mathrm{eff}}$ (early type) & \checkmark  & \checkmark  & - & \checkmark  & \checkmark  & - & \checkmark  & \checkmark  & \checkmark  \\ 
\mgi\ $8807$~\AA\ & $T_{\mathrm{eff}}$ & - & - & - & - & \checkmark  & - & - & \checkmark  & - \\

        \bottomrule
    \end{tabular} 
    \tablefoot{For tracers with multiple lines (e.g. doublets) or molecular bands, we average the feature importance values. The results are based on the feature importance evaluation of \settl\ (Figs. \ref{fig:fi_stl_m} and \ref{fig:fi_stl_kg}).}
\end{table*}

These results show that the features that our networks rely on to determine parameters vary depending on the input object. In particular, when changing from M- to K-type, important features change noticeably. For example, the \nai\ doublet $8183$, $8195$~\AA\ lines are essential features for networks to understand M-type stars, sensitive to all three stellar parameters, but for earlier type stars (K- and G-types), it is important only to determine gravity. Similarly, the \ki\ doublet lines are gravity-sensitive features for late-type stars but they are less essential for earlier types. In the case of \nai\ doublet $5890$, $5896$~\AA\ lines, on the other hand, they are more important for hot stars rather than for cold stars to determine gravity.

Please note that the feature-importance tests presented in this section indicate the features that affect the network's judgement which is based on the Phoenix models. Some of the important features (that are essential for the network) behave very similarly to our knowledge, but some do not. Above all, the behaviour of the \nai\ doublet $8183$, $8195$~\AA\ lines in the feature importance test agrees well with our knowledge. The \nai\ line, tracing the gravity~\citep{Riddick2007, Herczeg2014, Manara2017} and the temperature of late-type stars~\citep{Kirkpatrick1991, Allen1995, Riddick2007}, is also essential for networks to determine stellar parameters of late-type stars and gravity. Based on Table~\ref{table:fi_tracer}, we find that the R1 $8015$--$8130$~\AA, \ki\ doublet $7665$, $7699$~\AA, and \baii, \fei, and \cai\ blend 6497~\AA\ as well behave similarly to our knowledge. On the other hand, unlike our knowledge that the \caii\ IR triplet $8498$, $8542$, $8662$~\AA\ and \mgi\ $8807$~\AA\ trace the temperature~\citep{Kirkpatrick1991, Allen1995, Luhman2003, Manara2013, Herczeg2014}, the networks do not rely much on these lines to estimate the temperature.

In the feature-importance results of extinction, we showed the interesting results that there are particularly influential features although the extinction affects the overall shape of the spectrum, not the particular lines. One of the possible causes is the degeneracy between temperature and extinction. In our results, the features influential in determining the temperature tend to have high importance in extinction as well (e.g. \nai\ doublet $8183$, $8195$~\AA, VO band, and \ha). Due to the degeneracy between the two parameters, the over- or under-estimation of temperature can be compensated by over- or under-estimate of extinction. So, if the features important for temperature are scrambled, it can also affect the determination of extinction.
Another possible cause is that the network determines extinction based on correlations between multiple features. For example, if the network relies on the ratios between several features to the \ha, \ha\ may have relatively higher importance than others because scrambling the \ha\ affects all these ratios.

The feature importance only shows how much the error increases by scrambling a certain feature. Therefore, it is not easy to clearly understand the reasons for the error increment. Compared to the spectra of template stars, however, it is obvious that cINN captures important information from the point where absorption or emission exists. 
There are many features used to predict parameters besides the major features indicated in the figures or in the table, but the important point is that the most influential features are the same as the tracers we already know. This confirms that even though we do not exactly know how cINNs learn the hidden rules from the training data, what cINNs learned is very close to the physical knowledge we have.

\section{Simulation gap and the best network}
\label{sec:simulation_gap}
In sections \ref{sec:predictionSynth} and \ref{sec:predictionTpl}, we showed that, for the synthetic models, our cINNs predict stellar parameters perfectly and for the template stars, network predictions are in good agreement with literature values within a 5 to 10 per cent error. The difference between literature values and network predictions slightly varies depending on the characteristics of the template stars. In sections \ref{sec:ResimSynth} and \ref{sec:ResimTemplates}, we confirmed that resimulation of the spectrum based on the network prediction 
well restored the original input spectrum. This means that the network successfully finds the most suitable model that satisfies the given observational data, as the network is designed to. In other words, the very good resimulation results indicate that cINNs provided us with the best results within the physics it has learned.

Interestingly, the resimulated spectrum based on the network prediction is closer to the original input spectrum than the resimulated spectrum based on the literature values for template stars (see Fig.~\ref{fig:ResimTemplates_Spectrum_ID_SO797} and Table~\ref{tab:ResimTemplatesRMSEs}), despite the discrepancy between the network prediction and literature value. This can be considered to be one of the following two cases.
One is because there is a simulation gap, i.e. a gap between the physics within training data (i.e. the Phoenix atmosphere models), and the physics of the real world. The other is because of misclassification, meaning the literature value used as a reference in this paper is inaccurate. In the former case, no matter how perfectly trained the network is in terms of machine learning, it encounters inherent limitations. The simulation gap can be improved if we use better training data.

The three Phoenix libraries used in this paper reflect lots of important physics and characteristics of stellar atmosphere, but, of course, do not perfectly reflect reality. Therefore, we suspect that it is because of the simulation gap that the parameter predictions differ from the literature values even though the resimulation results are almost perfect. In this section, we will introduce how we can quantify the simulation gap using the trained cINN and determine how large the gap is between the Phoenix models and reality. Finally, we will draw comprehensive conclusions about the performance and usage of our cINNs.

\subsection{Quantifying simulation gap}
As explained in section~\ref{sec:cinn}, cINN consists of the main network that connects parameters (\textbf{x}) and latent variables (\textbf{z}) and conditioning network ($h$) that transforms the input observation (\textbf{y}) to the useful representative (i.e. condition, \textbf{c}). Both are trained together, and the conditioning network in this paper compresses 2930 features ($y_1, \ldots, y_{2930}$) included in one spectrum into 256 conditions ($c_1, \ldots, c_{256}$). If the condition of the real observational data passed through the conditioning network ($\textbf{c}_{\mathrm{obs}}$) follows the same probability distribution as the condition of the training data  ($\textbf{c}_{\mathrm{train}}$) this means there is no simulation gap. Because the conditioning network extracts only important features from the spectrum. 

However, unlike the latent variables set up to follow a prescribed distribution (i.e. a standard normal distribution), the distribution of conditions does not follow a certain known distribution.
Therefore, we build a network ($k$) that transforms the distribution of conditions ($p(\textbf{c})$) into a prescribed probability distribution. The $k$ network based on the cINN architecture is described as $k(\textbf{c}) = \textbf{s}$, and the output \textbf{s} is trained to follow a standard normal distribution. By definition of the cINN architecture, the dimensions of \textbf{c} and \textbf{s} are the same.

Using the conditioning network $h$ and transformation network $k$, we check the simulation gap between the Phoenix models and template stars by comparing the distribution of the transformed condition of template stars $k(h(\textbf{y}_{\mathrm{tpl}})) =  \textbf{s}_{\mathrm{tpl}}$ with the distribution of transformed condition of the training data $\textbf{s}_{\mathrm{train}}$ which follows a known distribution. We evaluate the simulation gap based on the $R^{2}$ score between two probability distributions, $p(\textbf{s}_{\mathrm{train}})$ and $p(\textbf{s}_{\mathrm{tpl}})$. The bigger the $R^{2}$ value, the smaller the simulation gap.

\subsection{Simulation gap}

\begin{figure}
	\includegraphics[width=1\columnwidth]{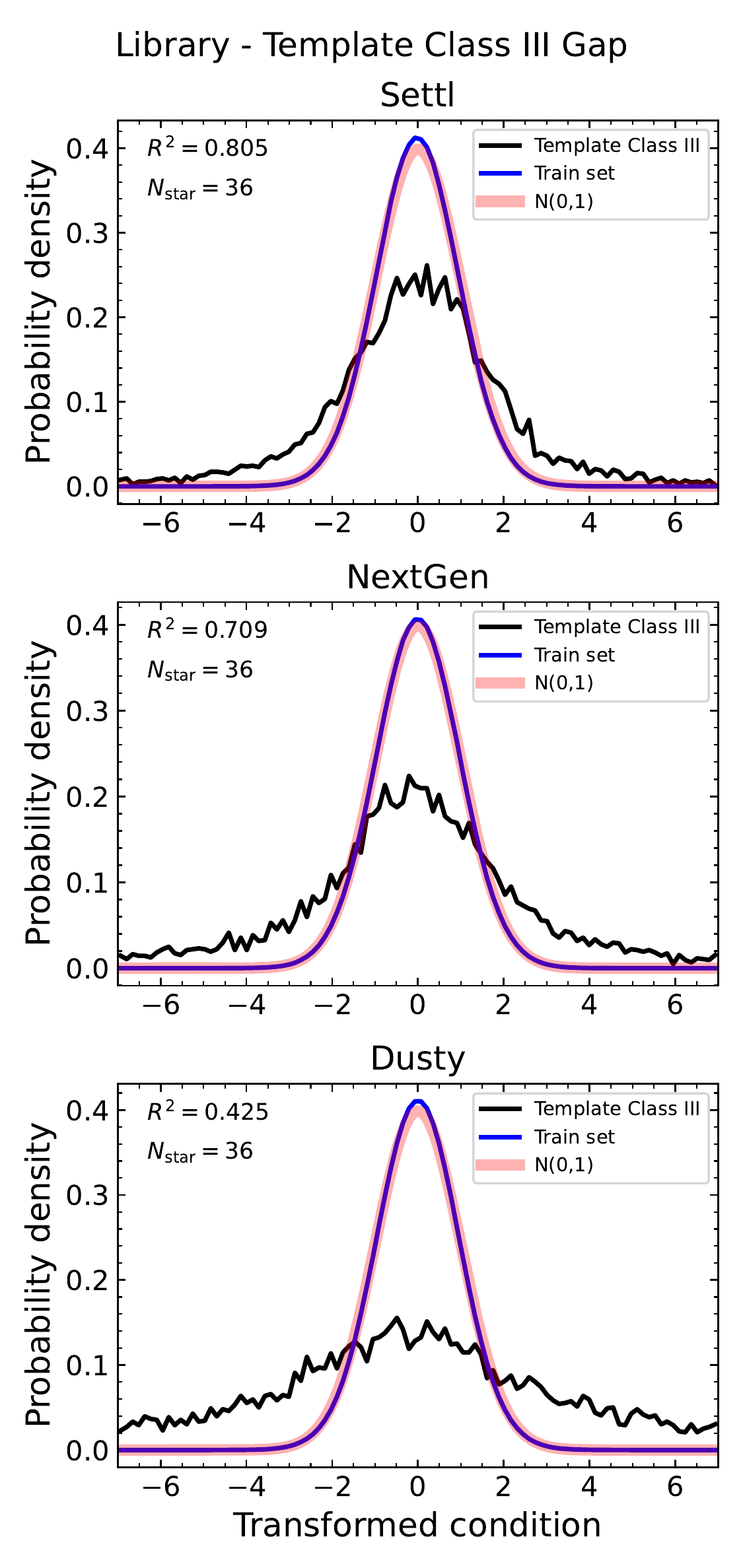}
    \caption{Probability distributions of transformed conditions of the training data (blue) and template stars (black) for three networks. The gap between the blue and black lines means the gap between the Phoenix model and the template spectrum. The $R^{2}$ value between the blue and black line and the number of template stars used are presented in the upper left corner of each panel.
     } \label{fig:sim_gap} 
\end{figure}

\begin{figure*}
	\includegraphics[width=2\columnwidth]{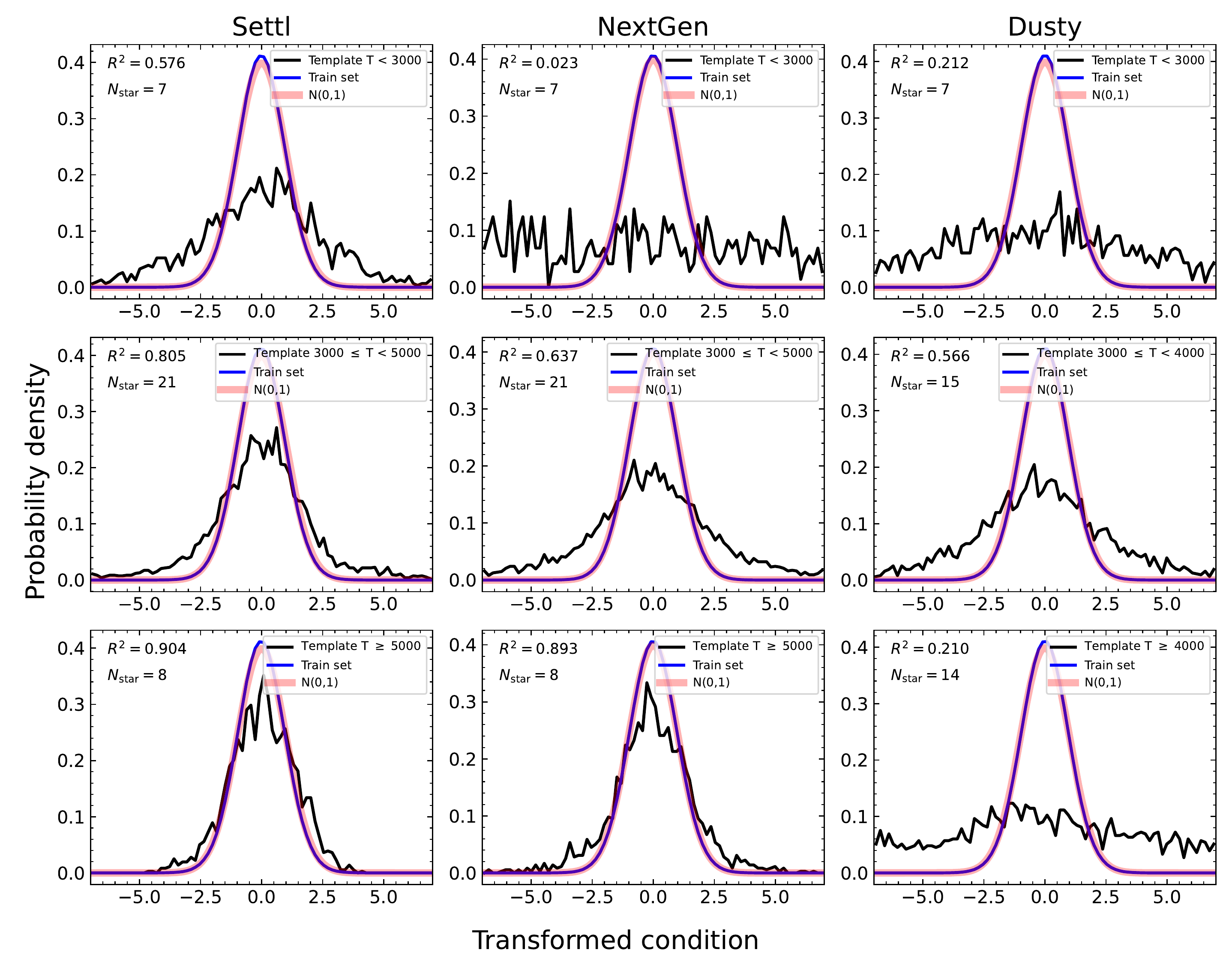}
    \caption{Probability distributions of transformed conditions of the training data and template stars. Each column represents three networks (\settl, \nextgen, and \dust) and each row represents the group of template stars depending on their temperature ($T^{\mathrm{lit}}_{\mathrm{eff}}$). Colour codes are the same as in Fig.~\ref{fig:sim_gap}.
     } \label{fig:sim_gap_T} 
\end{figure*}

We trained transformation networks ($k$) for each cINNs (\settl, \nextgen, and \dust) and compare the probability distribution of the transformed conditions of the training data and template stars. Figure~\ref{fig:sim_gap} shows that the distribution of the training data (blue line) well follows the prescribed standard normal distribution (pink line) but the distribution of template stars (black) differs from that of the training data. There are 256 condition components for each star, but we present all these components in one distribution. The $R^{2}$ scores for all template stars are 0.805, 0.709, and 0.425 for Settl, NextGen, and Dusty, respectively. The dusty model seems to have the widest simulation gap, but we need to consider that \dust\ has a narrower training range than the parameter space of the template stars. 

As the performance of the cINN varies depending on the temperature of the template star, we divided the stars into three groups based on the prediction performance of the networks shown in section~\ref{sec:predictionTpl} (see Figs.~\ref{fig:tpl_dt_comp} and \ref{fig:tpl_cerr_comp}). 
For example, \settl\ and \nextgen\ predicted parameters with good agreement with literature values, especially for stars with temperatures between $\sim$3000~K and $\sim$5000~K. So we divided the stars into 3 groups based on 3000~K and 5000~K for \settl\ and \nextgen. In the case of \dust, due to the temperature upper limit of 4000~K for the Dusty training set, we divided groups based on 3000~K and 4000~K.

In the case of Settl and NextGen libraries (Fig.~\ref{fig:sim_gap_T}), the earlier the spectral type, the smaller the gap and Settl has a smaller gap than NextGen in the overall temperature range. While the simulation gap is small for hot stars above 3000~K, the gap is large for later-type stars below 3000~K. In the case of NextGen, in particular, the simulation gap is very large for stars below 3000~K. In the case of Dusty, the simulation gap for the coldest group ($T < 3000$~K) is also very large and comparable to that for hot stars ($T > 4000$~K), out of the temperature range of the Dusty library.

The large gap for the lowest temperature group ($T < 3000$~K) is an obvious result because perfectly implementing the atmosphere of late-type stars through the simulation is a much more difficult task than the earlier type stars. For late-type stars, condensation of vapour is important but the relevant physical processes are complex, making it very difficult to produce a good atmosphere model. Thus, these results demonstrate the inherent limitations of modelling low-temperature stars. These results show that the degree of simulation gap varies with the characteristics of the star, just as the difference between the prediction of cINN and the literature value varies, as shown in section~\ref{sec:predictionTpl}.

Interestingly, both Settl and NextGen have the smallest simulation gaps for the early-type stars with temperatures above 5000~K. However, in Figs.~\ref{fig:tpl_dt_comp} and \ref{fig:tpl_cerr_comp}, the difference between the MAP prediction and the literature value of this group is slightly larger than that of the intermediate temperature group (3000--5000~K). The smallest simulation gap (Fig.~\ref{fig:sim_gap_T}) and good resimulation results better than the resimulation of literature values (Fig.~\ref{fig:ResimTemplates_Spectrum_ID_RXJ0445.8+1556} and Table~\ref{tab:ResimTemplatesRMSEs}) imply that MAP estimates of our networks for early-type stars above 5000~K are sufficiently reliable. Therefore, we suggest that the parameter estimations by our networks may be more accurate than the literature values for early-type stars above 5000~K.

\subsection{Best network}

It is clear that the simulation gap is large for late-type stars. Interestingly, however, our cINNs nevertheless predict the temperature and surface gravity well. First of all, all three networks had poor predictions of extinction for late-type stars below 3000~K. It is therefore very difficult for the network to estimate extinction accurately for stars in this temperature range, and the estimated extinction is not very reliable compared to the other two stellar parameters. However, \settl, \nextgen, and \dust\ estimated the temperature accurately with maximum errors of less than 10, 5, and 15 per cent, respectively, despite the large simulation gap. This is a sufficiently accurate prediction considering the temperature interval between 1 subclass of stellar spectral type (see Fig.~\ref{fig:tpl_dt_comp}). Using combined error in Fig.~\ref{fig:tpl_cerr_comp}, we demonstrated that \dust\ and \settl\ predict surface gravity and temperature accurately within 5 per cent for late-type stars as well as early-type stars, despite the simulation gap of late-type stars. This shows that our networks are still applicable to low-temperature stars despite the limitations of training data. In the case of \nextgen, its performance was relatively poor for low-temperature stars compared to the other two networks, which is explained by the large simulation gap shown in Fig.~\ref{fig:sim_gap_T}.

On the other hand, for earlier type stars with relatively small simulation gaps, the network performs more reliably. Except for one or two outliers, both \settl\ and \nextgen\ accurately predict temperature and gravity within a maximum error of 5 to 10 per cent. \nextgen\ tends to estimate extinction and temperature slightly higher than \settl. This seems that \nextgen\ is adopting a degenerate solution that satisfies the same input spectrum by increasing both extinction and temperature slightly. Overall, \settl, with the smallest simulation gap, shows the best performance among the three networks.

We conclude that \settl\ is the best network considering both parameter prediction performance and simulation gap. Against low-temperature stars (e.g. M-type stars), \dust\ also shows comparable performance to \settl. However, given that the stellar parameter coverage (i.e. temperature and gravity) of \settl\ is wider than that of \dust, \settl\ is more versatile and usable. Therefore, based on our overall results, we recommend using \settl\ when applying the network to real observations. The only limitation to be cautious of is the estimation of extinction. Regardless of the spectral type of the stars, cINN estimates temperature and gravity accurately, but it should be cautious of using estimated extinction when the estimated temperature is below 3000~K.

\section{Summary}
\label{sec:summary}
In this paper, we introduce a novel tool to estimate stellar parameters from the optical spectrum of an individual young, low-mass star. cINN is one of the deep learning architectures specialised in solving a degenerate inverse problem. The degenerate problem here means that, due to the inevitable information loss during the forward process from the physical system to observation, different physical systems are mapped onto similar or almost identical observations. Many of the major tasks in astrophysics are solving degenerate inverse problems like estimating physical properties from observations. In this work, we develop a cINN for young low-mass stars to efficiently diagnose their optical spectra and estimate stellar parameters such as effective temperature, surface gravity, and extinction.

cINN adopts a supervised learning approach, meaning that the network is trained on the database consisting of numerous well-labelled data sets of physical parameters and observations. However, it is difficult to collect a sufficient number of well-interpreted observations in real. Therefore, we use synthetic observations instead to generate enough training data. In this work, we utilise three Phoenix stellar atmosphere libraries (i.e. Settl, NextGen, and Dusty) to produce the database for training and evaluation of the network. Interpolating the spectrum on the temperature -- gravity space and adding the extinction effect on the synthetic spectra, we produce a database for each Phoenix library consisting of 65,536 synthetic models. To produce databases, we randomly sampled three parameters from the given parameter ranges. Settl and NextGen databases cover the temperature range of 2600--7000~K and $\log(g/\mathrm{cm\,s}^{-2})$ range of 2.5--5. In the case of the Dusty database, it covers the temperature of 2600--4000~K and $\log(g/\mathrm{cm\,s}^{-2})$ of 3--5. All three databases have extinction values within 0--10 mag. 
Then, we build and train cINNs using each database but only use 80\%\ of the synthetic models in the database to train the network and remain the rest for evaluation. In this paper, we present three cINNs that learned about different Phoenix atmosphere models: \settl, \nextgen, and \dust.

We validated the performance of our cINNs in various methods. Our main results are the following:

\begin{enumerate}
    \item All three networks provide perfect predictions on the test set with the RMSE of less than 0.01 dex for all three parameters, demonstrating that the cINNs are well-trained. Additionally, we resimulate the spectrum using the parameters estimated by the network using our interpolation method and compare it with the original input spectrum. The resimulated spectra perfectly match the input spectra of the test models with RMSE of about $10^{-7}$. These results prove that our three cINNs perfectly learned the hidden rules in each training data.
    
    \item To test the performance on the real observational data, we analyse 36 Class III template stars well-interpreted by \cite{Manara2013, Manara2017, Stelzer2013} with our cINNs. We demonstrate that stellar parameters estimated by our cINNs are in good agreement with the literature values. 
    
    \item Each network has a slightly different error depending on the temperature of the given star. \settl\ works especially well for M6.5 -- K1.0 (2935 -- 5000~K) stars and \nextgen\ works well for M4.5 -- K1.0 (3200 -- 5000~K) stars. \dust\ works well for M5.5 -- M0.0 (3060 -- 4000~K) stars. Given that the temperature upper limit of Dusty training data is 4000~K, \dust\ works well for stars within its training range. For stars in other temperature ranges, three networks perform well with an error of less than 10 per cent.
    
    \item The most difficult parameter for cINNs to predict is the extinction of cold stars with temperatures less than 3200~K. All three networks tend to estimate extinction higher than the literature value for cold stars. However, cINNs estimate extinction well for hot stars with temperatures above 3200~K.

    \item We resimulate spectra based on cINN estimations and literature values and compare them with the original input spectrum. Interestingly, most of the resimulated spectra based on cINN estimations are closer to the input spectra than the resimulated spectra derived from literature values. This implies that our cINNs well understand the physics in each Phoenix library and are able to find the best-fitting Phoenix model (i.e. parameters) for the given observation.
    
    \item Results that the resimulations are perfect even though the prediction of the network is slightly different from the literature can be explained by a gap between the Phoenix model and reality, so-called the simulation gap. We quantify the simulation gap between each library and template stars using the conditioning networks included in our cINNs. We confirm that the simulation gaps are relatively large for cold stars below 3000~K where the cINNs have difficulty estimating extinction. We confirm that the simulation gap is small for hot stars where cINNs predict parameters well. 
    
    \item The overall results imply that although there is an obvious gap between the Phoenix model and reality, especially for cold stars below 3000~K, our networks can nonetheless provide reliable predictions for all stars within 5--10 per cent error, especially for temperature and gravity. Extinction estimated by cINN is also reliable unless the estimated temperature is less than 3200~K.

    \item We investigate which parts of the spectrum cINN relies mostly upon to predict stellar parameters and compare the important features with typically used stellar parameter tracers. We find that cINN relies on different features depending on the physical parameters and on the input observations (e.g. spectral types). We confirm that the major features are equivalent to the typically used tracers such as \ha\ $6563$~\AA\ and \nai\ doublet $8183$, $8195$~\AA. 

\end{enumerate}

Our overall results show that our cINNs present reliable enough performance applicable to real observational data. Among the three networks introduced in this paper, we recommend \settl\ trained on the Settl library as the best network because of its remarkable performance and versatility on the parameter space.

\begin{acknowledgements}

This work was partly supported by European Union’s Horizon 2020 research and innovation program
and the European Research Council via the ERC Synergy Grant ``ECOGAL'' (project ID 855130),  and the Marie Sklodowska-Curie grant DUSTBUSTERS (project No 823823),  by the Deutsche Forschungsgemeinschaft (DFG) via the Collaborative Research Center ``The Milky Way System''  (SFB 881 -- funding ID 138713538 -- subprojects A1, B1, B2 and B8), by the Heidelberg Cluster of Excellence (EXC 2181 - 390900948) ``STRUCTURES'', funded by the German Excellence Strategy, and by the German Ministry for Economic Affairs and Climate Action in project ``MAINN'' (funding ID 50OO2206). We also thank for computing resources provided by the Ministry of Science, Research and the Arts (MWK) of the State of Baden-W\"{u}rttemberg through bwHPC and DFG through grant INST 35/1134-1 FUGG and for data storage at SDS@hd through grant INST 35/1314-1 FUGG.

\end{acknowledgements}

\bibliographystyle{aa}


\begin{appendix} 

\section{Supplemental materials}
In this appendix, we present supplementary figures and table mentioned in our main results (sections~\ref{sec:validation}--\ref{sec:featureimportance}).

\subsection{Prediction performance}

\begin{figure*}
	\includegraphics[width=2\columnwidth]{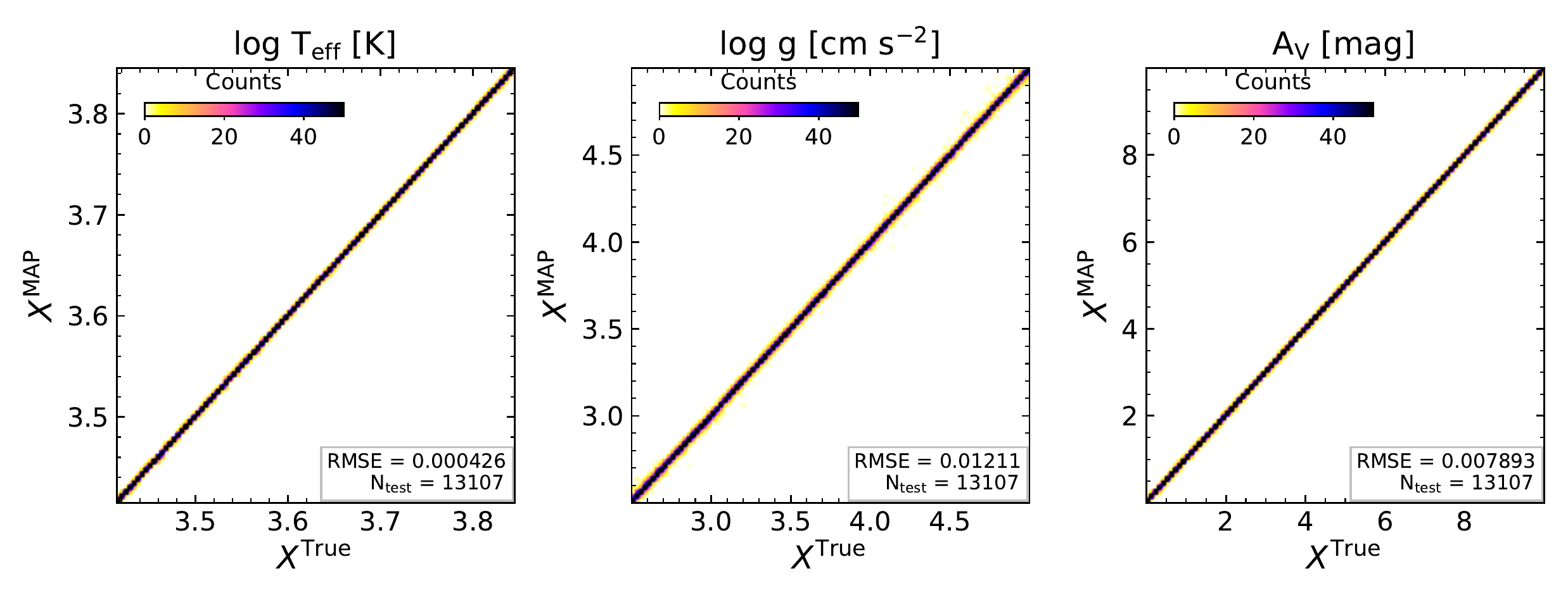}
    \caption{2-dimensional histograms comparing the MAP values estimated by the \settl\ and the true values for the entire test models of the Settl database. The colours indicate the number of models at each point in the 2D histograms.
    In the lower right corner, we present the root mean square error (RMSE) and the number of test models used (N$_{\mathrm{test}}$).
     } \label{fig:settl_test_map} 
\end{figure*}

We evaluate the performance of three networks (\settl, \nextgen, and \dust) on 13,107 synthetic test models drawn from the corresponding database by comparing the MAP predictions from the network and the true values of the models. Here, we present the result of \settl\ in Fig.~\ref{fig:settl_test_map} as a representative because the other two networks (\nextgen\ and \dust) also show very similar results. The figure shows that the network estimates all three parameters perfectly with very small RMSEs.

\clearpage



\subsection{Resimulation}
We validate the cINN predictions on both the synthetic test data (Sect.~\ref{sec:ResimSynth}) and real template spectra (Sect.~\ref{sec:ResimTemplates}) by resimulating the spectra corresponding to the MAP estimates with our spectral library interpolator (Sect.~\ref{sec:spectral_libraries}) and comparing the result to the respective input spectra. 

\begin{figure*}
	\includegraphics[width=\textwidth]{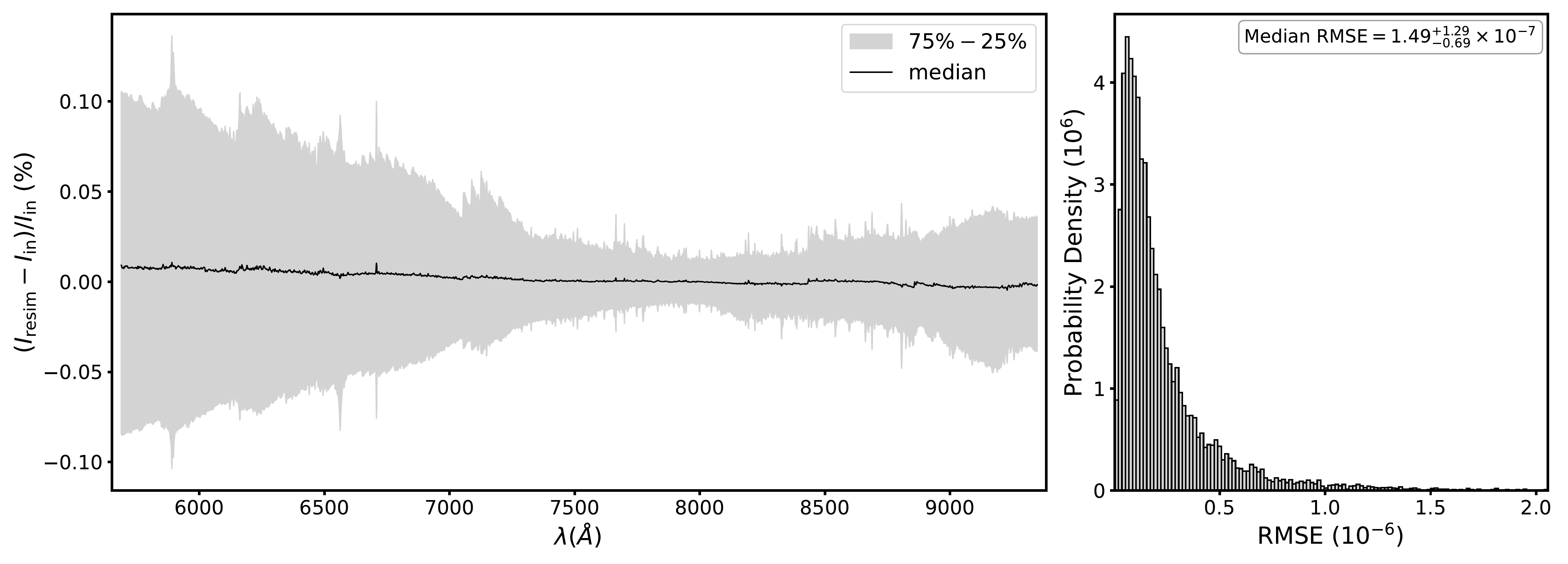}
    \caption{Resimulation results of \nextgen\ for the entire synthetic spectra in the test set.
    Left: Median relative error across the wavelength range of the resimulated spectra based on the MAP predictions of the cINN trained on the NextGen models averaged over the 13,107 synthetic spectra in the test set. Here the grey envelope indicates the interquantile range between the $25\%$ and $75\%$ quantiles. Right: Histogram of the RMSEs of the 13,107 resimulated spectra. The mean resimulation RMSE across the test set is $2.28 \pm 2.48 \times 10^{-7}$. }
    \label{fig:ResimSynth_NextGen}
\end{figure*}

\begin{figure*}
	\includegraphics[width=\textwidth]{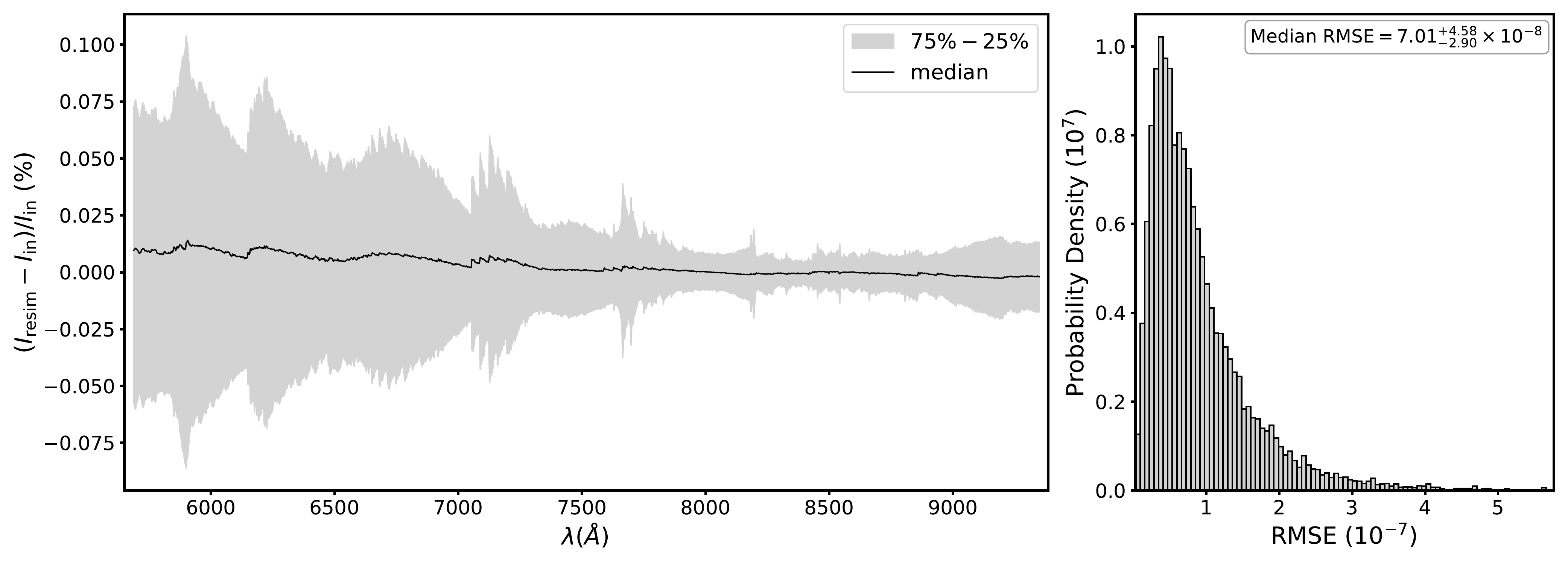}
    \caption{Resimulation results of \dust\ for the entire synthetic spectra in the test set.
    Left: Median relative error across the wavelength range of the resimulated spectra based on the MAP predictions of the cINN trained on the Dusty models averaged over the 13,107 synthetic spectra in the test set. Here the grey envelope indicates the interquantile range between the $25\%$ and $75\%$ quantiles. Right: Histogram of the RMSEs of the 13,107 resimulated spectra. The mean resimulation RMSE across the test set is $9.01 \pm 7.34 \times 10^{-8}$. }
    \label{fig:ResimSynth_Dusty}
\end{figure*}

Analogous to Fig.~\ref{fig:ResimSynth_Settl}, Figs.~\ref{fig:ResimSynth_NextGen} and \ref{fig:ResimSynth_Dusty} show the median relative error of the resimulated spectra (left panel) and distributions of the RMSEs (right panel) for the 13,107 synthetic test spectra when evaluated with the cINN models trained on NextGen and Dusty, respectively.

Table~\ref{tab:ResimTemplatesRMSEs} provides a summary of the resimulation results for the cINN predictions on the Class III template spectra (see Sect.~\ref{sec:ResimTemplates} and also Tables~\ref{tab:class_III_templates} and \ref{tab:SummaryTemplateMAP}). Here we list the RMSEs and $R^2$ scores of the resimulated spectra with respect to the corresponding input spectra for the resimulation based on the literature and cINN-predicted parameters for all three spectral libraries.

\begin{figure*}
    \centering
    \includegraphics[width=2\columnwidth]{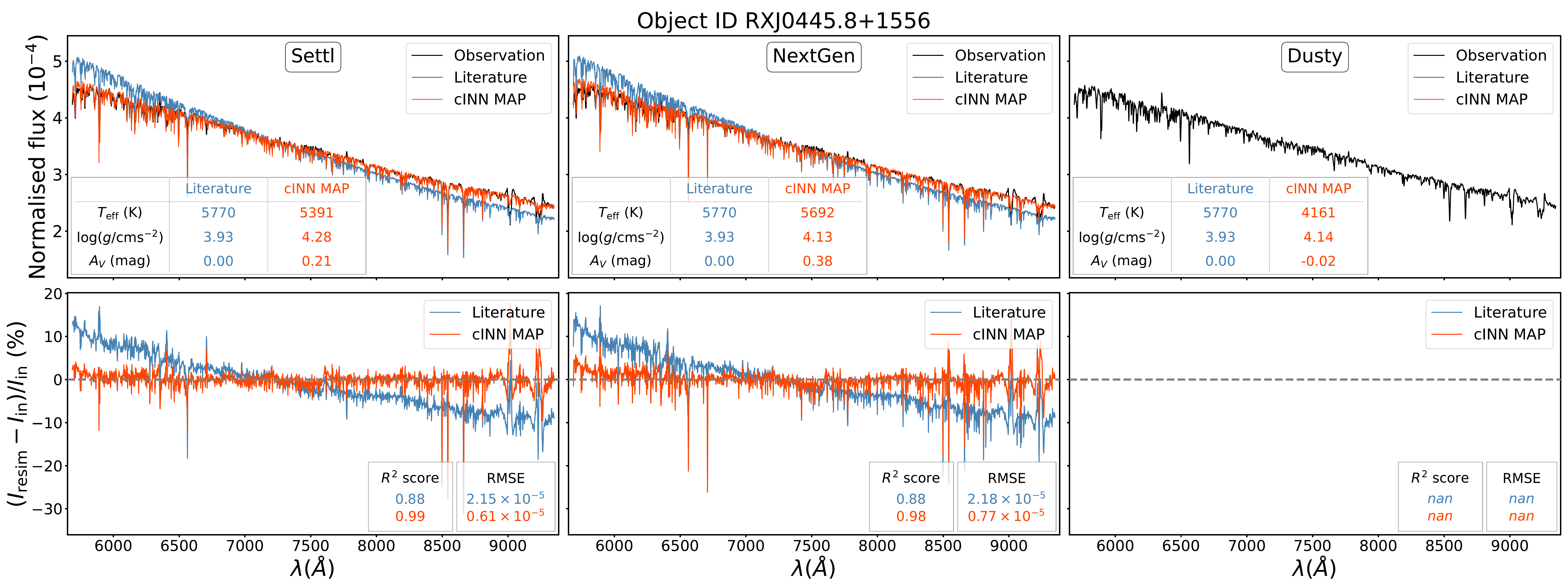}
    \caption{Resimulation results for Class III star RXJ0445.8+1556. Same as Fig.~\ref{fig:ResimTemplates_Spectrum_ID_SO797}.}
    \label{fig:ResimTemplates_Spectrum_ID_RXJ0445.8+1556}
\end{figure*}

\begin{figure*}
    \centering
    \includegraphics[width=2\columnwidth]{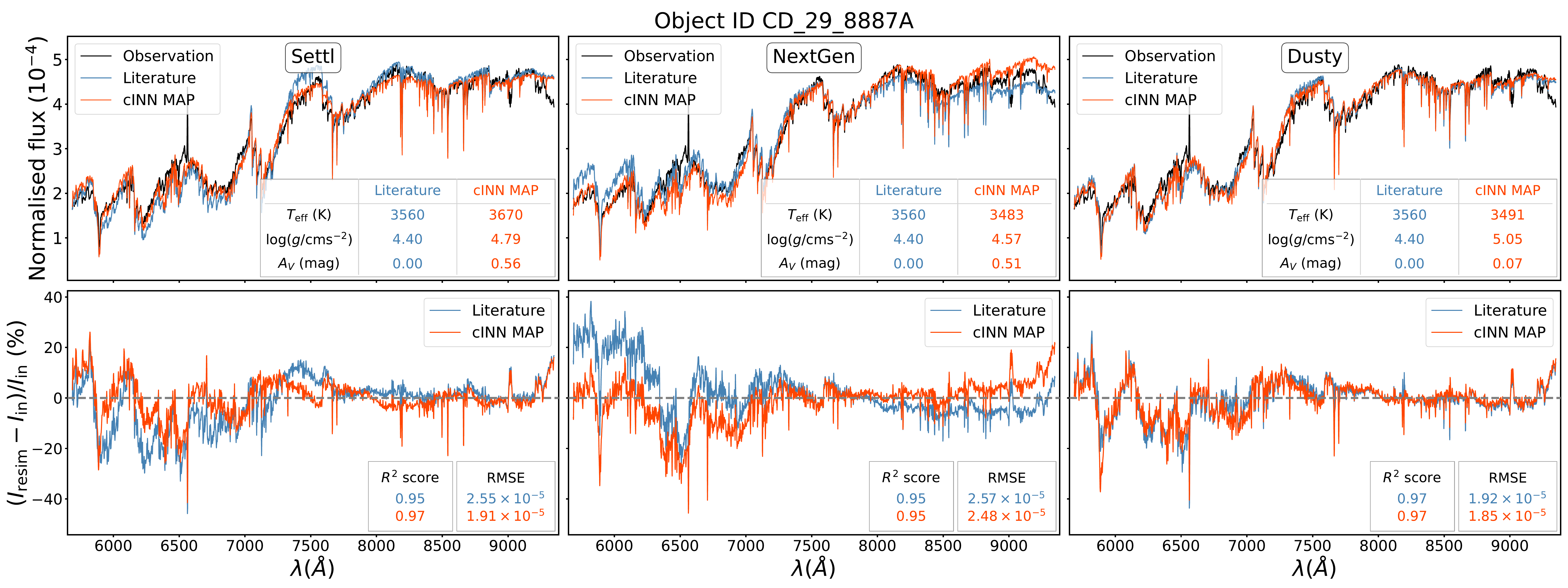}
    \caption{Resimulation results for Class III star CD\_29\_8887A. Same as Fig.~\ref{fig:ResimTemplates_Spectrum_ID_SO797}.}
    \label{fig:ResimTemplates_Spectrum_ID_CD_29_8887A}
\end{figure*}

Figures~\ref{fig:ResimTemplates_Spectrum_ID_RXJ0445.8+1556} and \ref{fig:ResimTemplates_Spectrum_ID_CD_29_8887A} provide additional examples of the resimulation results, comparing the resimulated spectra to the input spectra and the outcomes between the three libraries, analogous to Fig.~\ref{fig:ResimTemplates_Spectrum_ID_SO797}. In particular, these two Figs. show examples, where the resimulated spectra based on the cINN MAP estimates seem to match the input spectra notably better than the respective resimulation outcome based on the literature properties of the given Class III templates.

\begin{figure*}
    \centering
    \includegraphics[width = 0.99\textwidth]{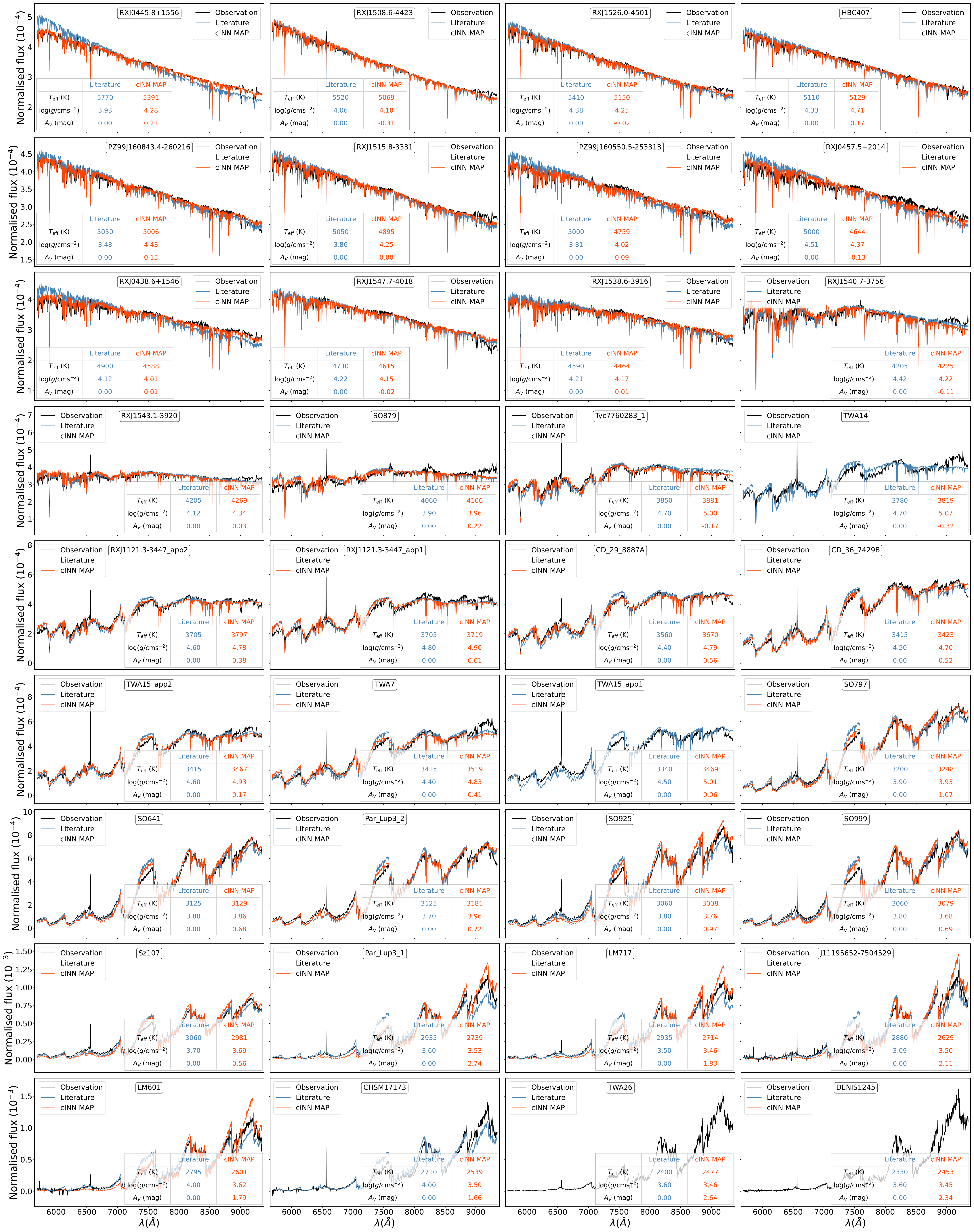}
    \caption{Resimulation results for all Class III templates for the cINN trained on the Settl library. In each panel, the black curve indicates the observed spectrum, while the red and blue curves correspond to the spectra resimulated based on the cINN MAP estimates and literature properties, respectively. The latter values are summarised in the table in each panel. Note that if either the red or blue or both curves are missing that the corresponding set of parameters could not be resimulated. For the RMSEs and $R^2$ scores of the resimulated spectra see Table~\ref{tab:ResimTemplatesRMSEs}.}
    \label{fig:ResimSettlAllSpectra}
\end{figure*}

Lastly, Fig.~\ref{fig:ResimSettlAllSpectra} provides an overview of the resimulation results for all Class III template spectra for the cINN trained on the Settl library, corresponding to the top left panels in Figs.~\ref{fig:ResimTemplates_Spectrum_ID_SO797}, \ref{fig:ResimTemplates_Spectrum_ID_RXJ0445.8+1556} and \ref{fig:ResimTemplates_Spectrum_ID_CD_29_8887A}.

\begin{table*}
    \centering
    \caption{Summary of the resimulation test for the literature values and cINN MAP predictions for the three different spectral libraries, listing the RMSEs and $R^2$ scores of the resimulated spectra.}
    \resizebox{\textwidth}{!}{
    \begin{tabular}{lccccccccc}
         \toprule
         & \multicolumn{9}{c}{Resimulation RMSE ($\times 10^{-5}$) / $R^2$ Score} \\
         \cmidrule(rl){2-10} 
         & \multicolumn{3}{c}{Settl} & \multicolumn{3}{c}{NextGen} & \multicolumn{3}{c}{Dusty} \\
         \cmidrule(rl){2-4} \cmidrule(rl){5-7}  \cmidrule(rl){8-10}  
         Object Name & Literature & cINN & Comment & Literature & cINN & Comment & Literature & cINN & Comment \\
         \midrule
         RXJ0445.8+1556         & 2.15 / 0.88 & 0.61 / 0.99 & -   & 2.18 / 0.88 & 0.77 / 0.98 & -   & - / - & - / - & $T_\mathrm{eff} > 4000$ K\\
         RXJ1508.6-4423         & 0.95 / 0.98 & 0.93 / 0.98 & -   & 1.08 / 0.98 & 1.05 / 0.98 & -   & - / - & - / - & $T_\mathrm{eff} > 4000$ K\\
         RXJ1526.0-4501         & 1.14 / 0.97 & 0.66 / 0.99 & -   & 1.20 / 0.96 & 0.77 / 0.98 & -   & - / - & - / - & $T_\mathrm{eff} > 4000$ K\\
         HBC407                 & 1.00 / 0.97 & 0.68 / 0.99 & -   & 1.16 / 0.96 & 0.91 / 0.97 & -   & - / - & - / - & $T_\mathrm{eff} > 4000$ K\\
         PZ99J160843.4-260216   & 1.08 / 0.96 & 0.82 / 0.98 & -   & 1.19 / 0.95 & 0.93 / 0.97 & -   & - / - & - / - & $T_\mathrm{eff} > 4000$ K\\
         RXJ1515.8-3331         & 1.28 / 0.94 & 0.85 / 0.97 & -   & 1.41 / 0.92 & 0.92 / 0.97 & -   & - / - & - / - & $T_\mathrm{eff} > 4000$ K\\
         PZ99J160550.5-253313   & 1.50 / 0.90 & 0.73 / 0.98 & -   & 1.64 / 0.88 & 0.93 / 0.96 & -   & - / - & - / - & $T_\mathrm{eff} > 4000$ K\\
         RXJ0457.5+2014         & 1.98 / 0.78 & 1.23 / 0.92 & -   & 2.07 / 0.76 & 1.25 / 0.91 & -   & - / - & - / - & $T_\mathrm{eff} > 4000$ K\\
         RXJ0438.6+1546         & 2.09 / 0.70 & 0.89 / 0.95 & -   & 2.19 / 0.67 & 1.02 / 0.93 & -   & - / - & - / - & $T_\mathrm{eff} > 4000$ K\\
         RXJ1547.7-4018         & 0.90 / 0.97 & 0.95 / 0.96 & -   & 1.08 / 0.95 & 1.11 / 0.95 & -   & - / - & - / - & $T_\mathrm{eff} > 4000$ K\\
         RXJ1538.6-3916         & 1.05 / 0.93 & 0.92 / 0.94 & -   & 1.24 / 0.90 & 1.20 / 0.91 & -   & - / - & - / - & $T_\mathrm{eff} > 4000$ K\\
         RXJ1540.7-3756         & 1.42 / 0.56 & 1.61 / 0.44 & -   & 1.56 / 0.48 & 1.54 / 0.49 & -   & - / - & - / - & $T_\mathrm{eff} > 4000$ K\\
         RXJ1543.1-3920         & 1.48 / 0.42 & 1.63 / 0.30 & -   & 1.72 / 0.22 & 1.48 / 0.42 & -   & - / - & - / - & $T_\mathrm{eff} > 4000$ K\\
         SO879                  & 2.66 / 0.53 & 2.44 / 0.61 & -   & 3.02 / 0.39 & 2.18 / 0.68 & -   & - / - & 2.08 / 0.71 & -\\
         Tyc7760283\_1          & 2.56 / 0.73 & 1.88 / 0.85 & -   & 1.99 / 0.84 & 1.95 / 0.84 & -   & 1.93 / 0.85 & 1.89 / 0.85 & $5 < \log(g) < 5.5$\\
         TWA14                  & 3.04 / 0.83 & -    / -    & $\log(g) > 5$   & 3.28 / 0.80 & 2.74 / 0.86 & -   & 2.96 / 0.84 & 3.07 / 0.82 & $5 < \log(g) < 5.5$\\
         RXJ1121.3-3447\_app2   & 2.19 / 0.93 & 1.88 / 0.95 & -   & 2.45 / 0.91 & 2.15 / 0.93 & -   & 1.86 / 0.95 & 1.80 / 0.95 & $5 < \log(g) < 5.5$\\
         RXJ1121.3-3447\_app1   & 2.69 / 0.92 & 2.84 / 0.91 & -   & 3.60 / 0.85 & 2.44 / 0.93 & -   & 2.87 / 0.91 & 2.42 / 0.93 & $5 < \log(g) < 5.5$\\
         CD\_29\_8887A          & 2.55 / 0.95 & 1.91 / 0.97 & -   & 2.57 / 0.95 & 2.48 / 0.95 & -   & 1.92 / 0.97 & 1.85 / 0.97 & $5 < \log(g) < 5.5$\\
         CD\_36\_7429B          & 2.70 / 0.97 & 2.30 / 0.98 & -   & 4.90 / 0.91 & 2.57 / 0.97 & -   & 3.26 / 0.96 & 2.26 / 0.98 & - \\
         TWA15\_app2            & 2.98 / 0.96 & 3.04 / 0.96 & -   & 4.04 / 0.92 & 2.59 / 0.97 & -   & 2.93 / 0.96 & 2.57 / 0.97 & $5 < \log(g) < 5.5$\\
         TWA7                   & 3.45 / 0.95 & 3.62 / 0.94 & -   & 4.53 / 0.91 & 2.66 / 0.97 & -   & 3.36 / 0.95 & 2.76 / 0.97 & - \\
         TWA15\_app1            & 3.95 / 0.93 & -    / -    & $\log(g) > 5$   & 3.26 / 0.95 & 2.95 / 0.96 & -   & 3.01 / 0.96 & 2.96 / 0.96 & $5 < \log(g) < 5.5$\\
         SO797                  & 3.77 / 0.97 & 2.70 / 0.98 & -   & 6.35 / 0.92 & 2.47 / 0.99 & -   & 4.63 / 0.96 & 2.27 / 0.99 & - \\
         SO641                  & 3.83 / 0.97 & 3.15 / 0.98 & -   & 6.37 / 0.92 & 2.62 / 0.99 & -   & 4.76 / 0.96 & 2.63 / 0.99 & - \\
         Par\_Lup3\_2           & 3.68 / 0.97 & 3.03 / 0.98 & -   & 4.74 / 0.95 & 2.86 / 0.98 & -   & 3.31 / 0.98 & 2.76 / 0.98 & - \\
         SO925                  & 4.55 / 0.97 & 4.42 / 0.97 & -   & 7.28 / 0.91 & 3.06 / 0.98 & -   & 5.91 / 0.94 & 3.17 / 0.98 & - \\
         SO999                  & 4.20 / 0.97 & 3.90 / 0.97 & -   & 6.27 / 0.93 & 2.99 / 0.98 & -   & 5.00 / 0.96 & 3.10 / 0.98 & - \\
         Sz107                  & 4.44 / 0.97 & 4.85 / 0.96 & -   & 6.58 / 0.93 & 2.83 / 0.99 & -   & 5.32 / 0.95 & 3.11 / 0.98 & - \\
         Par\_Lup3\_1           & 8.90 / 0.92 & 5.64 / 0.97 & -   & 12.4 / 0.85 & -    / -    & $\log(g) < 2.5$   & 11.5 / 0.87 & 4.05 / 0.98 & - \\
         LM717                  & 7.08 / 0.95 & 5.77 / 0.96 & -   & 10.1 / 0.88 & -    / -    & $\log(g) < 2.5$   & 9.80 / 0.90 & -    / -    & $\log(g) < 3.0$   \\
         J11195652-7504529      & 7.49 / 0.95 & 6.73 / 0.96 & -   & 10.9 / 0.89 & -    / -    & $\log(g) < 2.5$   & 10.1 / 0.89 & -    / -    & $\log(g) < 3.0$   \\
         LM601                  & 7.76 / 0.94 & 7.26 / 0.95 & -   & 9.97 / 0.91 & -    / -    & $\log(g) < 2.5$   & 9.06 / 0.92 & -    / -    & $\log(g) < 3.0$   \\
         CHSM17173              & 8.65 / 0.94 & -    / -    & $T_\mathrm{eff} < 2700$ K   & 10.1 / 0.90 & -    / -    & $\log(g) < 2.5$   & 9.63 / 0.92 & -    / -    & $\log(g) < 3.0$   \\
         TWA26                  & -    / -    & -    / -    & $T_\mathrm{eff} < 2700$ K   & -    / -    & -    / -    & $\log(g) < 2.5$   & -    / -    & -    / -    & $T_\mathrm{eff} < 2700$ K   \\
         DENIS1245              & -    / -    & -    / -    & $T_\mathrm{eff} < 2700$ K   & -    / -    & -    / -    & $\log(g) < 2.5$   & -    / -    & -    / -    & $T_\mathrm{eff} < 2700$ K   \\
         \midrule
         Resimulated Spectra   & 34   & 31    & -    & 34   & 29   & -    & 20   & 17   & -    \\ 
         \bottomrule
    \end{tabular}}
    \label{tab:ResimTemplatesRMSEs}
    \tablefoot{The comment column indicates the reason why the cINN prediction could not be resimulated. Note that for SO879 the cINN prediction can be resimulated with the Dusty library despite the literature temperature being above 4000 K, because the cINN underestimates $T_\mathrm{eff}$ by 151~K here, thus falling into the Dusty temperature boundaries.}
\end{table*}

\FloatBarrier

\subsection{Feature importance}
\begin{figure*}
	\includegraphics[width=1\columnwidth]{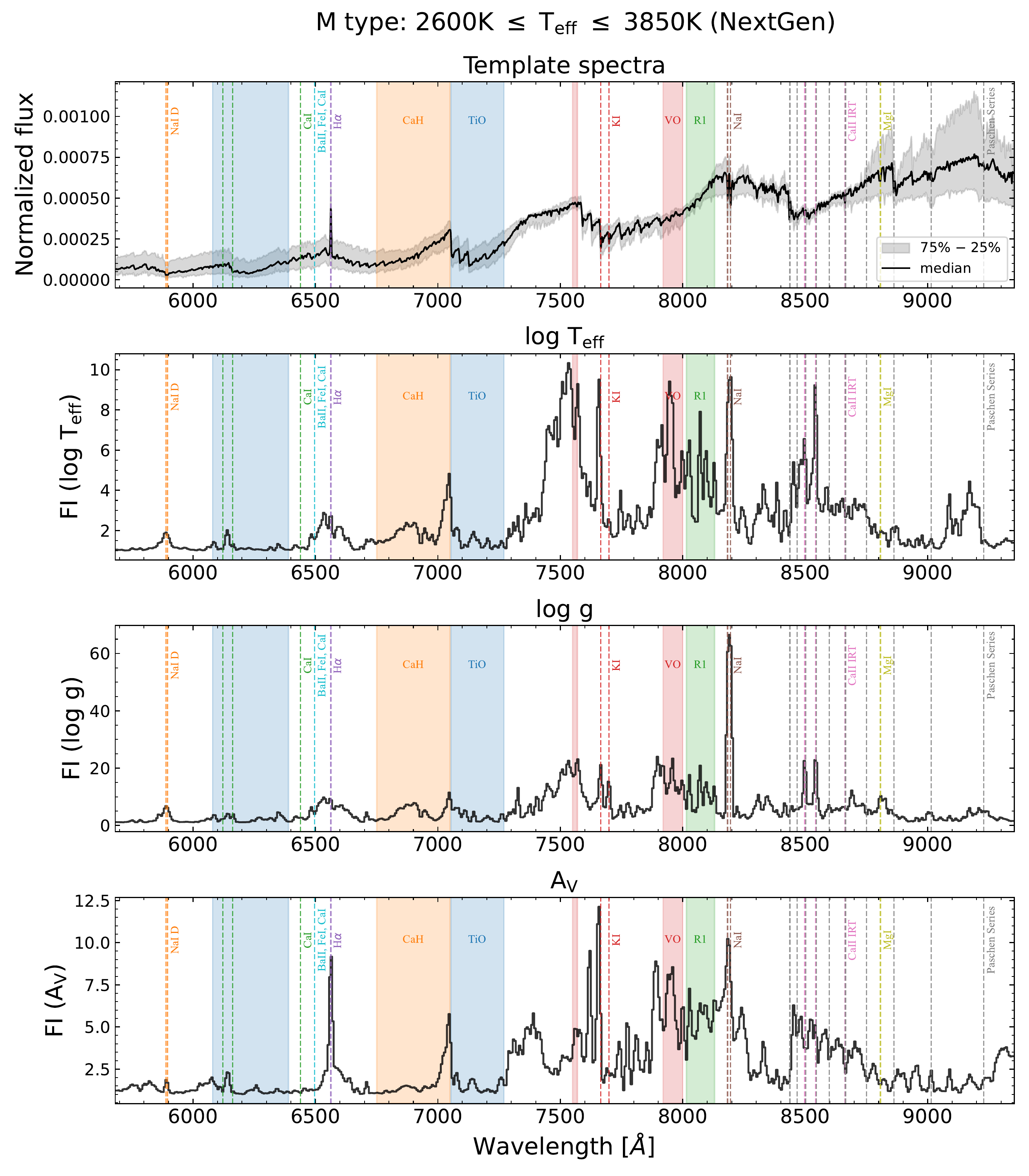}
	\includegraphics[width=1\columnwidth]{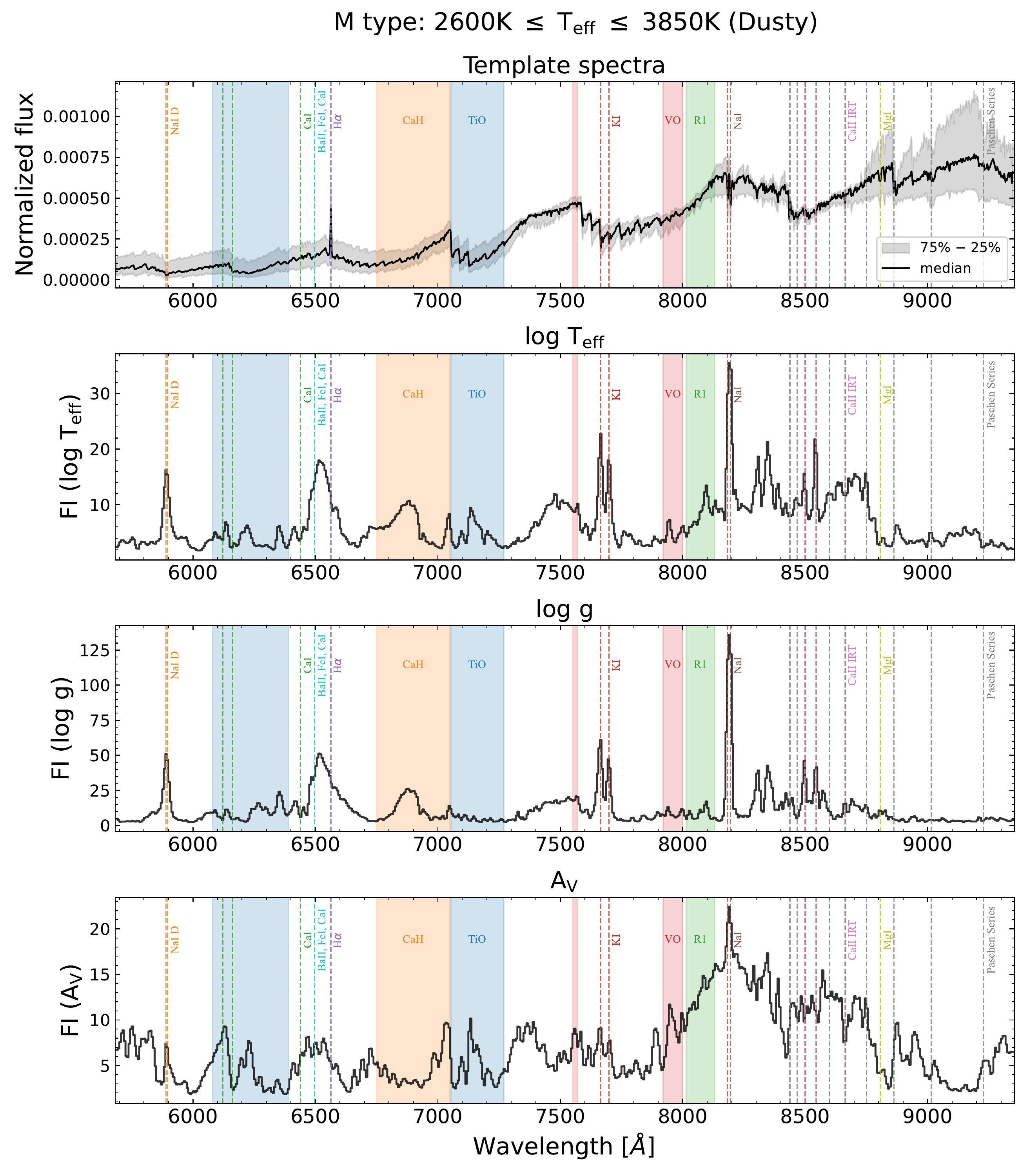}
    \caption{Feature importance evaluation for M-type synthetic models in the test set using \nextgen\ (left) and \dust\ (right), respectively. The first row shows the median flux of M-type Class III template stars. Lines and shades are the same as Fig.~\ref{fig:fi_stl_m}.
     }
     \label{fig:fi_ng_m}
\end{figure*}

We investigate the important feature where \nextgen\ and \dust\ rely mostly upon. We divided the synthetic observations into three groups depending on their spectral types (e.g. M-, K-, and G-types). We present the results of \nextgen\ and \dust\ for M-type stars in Fig.~\ref{fig:fi_ng_m}. We do not present the results of \nextgen\ for K- and G-type stars because overall results are similar to that of \settl\ presented in Fig.~\ref{fig:fi_stl_kg}.

\FloatBarrier

\end{appendix}

\end{document}